\documentclass[%
twocolumn,
nofootinbib,
 amsmath,amssymb,
 aps,
floatfix,
superscriptaddress,showpacs]{revtex4-2}

\usepackage[utf8]{inputenc}
\usepackage{graphicx}
\usepackage{dcolumn}
\usepackage{bm}
\usepackage{lineno}
\usepackage{float}
\usepackage{xspace}
\usepackage[section]{placeins}
\usepackage{xcolor}
\usepackage{gensymb}
\begin{document}

\newcommand{\sigreac}{$\sigma_{reac}$\xspace}
\newcommand\czer[1]{{\color{red}#1}}

\title{Comparison of Validation Methods of Simulations for
Final State Interactions in Hadron Production Experiments}
\author{S. Dytman}\affiliation{University of Pittsburgh, Department of Physics and Astronomy, Pittsburgh, PA 15260, USA}
\author{Y. Hayato}
\affiliation{University of Tokyo, ICRR, Kamioka, Japan}
\author{R. Raboanary}\affiliation{University of Antananarivo, Antananarivo, Madagascar} 
\author{J.T. Sobczyk}\affiliation{University of Wroc\l{}aw, Institute of Theoretical Physics, Wroc\l{}aw, Poland}
\author{J. Tena-Vidal}\affiliation{University of Liverpool, Liverpool, UK}
\author{N. Vololoniaina}\affiliation{University of Antananarivo, Antananarivo, Madagascar} 

\date{March, 2021}

\begin{abstract}
Neutrino cross section and oscillation measurements depend critically on modeling of hadronic final state interactions (FSI).  Often, this is one of the largest components of uncertainty in a measurement.  This is because of the difficulty in modeling strong interactions in nuclei in a consistent quantum-mechanical framework.  FSI models are most often validated using hadron-nucleus data which introduces further uncertainties.  The alternative is to use transparency data where the hadron starts propagating from inside the nucleus and the probability of interaction is measured as a function of hadron energy.  This work examines the relationship between the $\pi^+$ and proton total reaction cross section and transparency from a simulation viewpoint.
\end{abstract}

\maketitle

\section{Introduction} 




 

A firm knowledge of neutrino interactions is required to properly assess backgrounds and systematic errors in neutrino oscillation measurements~\cite{Abe:2018wpn,Acero:2019ksn,Abi:2020evt,Antonello:2015lea}.
Oscillation experiments use pion and proton detection to varying degrees.  Some detect these hadrons within the oscillation analysis~\cite{Antonello:2015lea}, but all depend on accurately separating events by interactions based on final state topology.
Recent oscillation articles from T2K~\cite{Abe:2018wpn} and NOvA~\cite{Acero:2019ksn} highlight this need.
A recent review~\cite{nustec-review} studies the various channels required to describe neutrino interactions and highlights the issues in the need for precise modeling.

Event generator codes simulate a large number of neutrino-nucleus interactions across a broad range of energies.
For proper detector modeling, the entire final state of particles is produced.
These codes are then used to produce Monte Carlo events for each experiment which are then used to identify event categories, deduce neutrino beam energy, assess background and estimate systematic errors.
The neutrino interacts with one of more nucleons through what is called the primary interaction.  
Hadrons are always emitted through this first interaction which is governed by the weak force; they then propagate through the residual nucleus according to final state interactions (FSI) modeling which is governed by the strong force.
At hadron kinetic energies characteristic for long baseline neutrino oscillation experiments (typically less than 1 GeV), the probability of hadron reinteraction is quite large. For pions produced mostly through the $\Delta(P_{33}(1232))$ excitation mechanism, $\sim 75\%$ of the pions interact for carbon and $\sim 80\%$ for argon. For protons the probability to reinteract depends on its kinetic energy and is of the order $\sim 40-50\%$.  Largely because of hadronic reinteraction, the energy transferred to nucleus in the primary interaction is shared by many particles knocked out of the nucleus.  Any measurement of the neutrino energy by measuring energy of all particles in the final state~\cite{Acero:2019ksn} must model FSI effects carefully. A large fraction of that energy is due to neutral particles such as neutrons and $\pi^0$ which are not easily visible in today's fine grained detectors~\cite{Elkins:2019vmy}.  Since FSI significantly modifies both the composition and energy/angle distributions of knocked-out hadrons, a proper description is a key component of each code.

One interesting issue is that experiments use a variety of codes for their simulations.  
These codes employ models of the same physics channels, but can use either different sources or different interpretations of the same source. 
As a result, measurements of neutrino oscillations can depend on these choices and the collaborations must properly assess the associated systematic uncertainties.
One important way to understand this situation is to produce comparisons of the codes against appropriate benchmark measurements. 
Many studies of this type have been presented in a series of NuInt workshops, e.g.~\cite{nuint6,Boyd:2009zz}.  Results from a recent broad study that allowed increased understanding of the issues were published in Ref.~\cite{tensions2016}.  
One major theme has been the use of event generator codes to examine the relationship between measurements that are not directly comparable.

FSI models are complex because hadrons propagate through the residual nucleus and can interact with any of the nucleons in it with high probability.
In addition, the struck nucleon may be interacting with another nucleon at the same time (called medium corrections or nucleon-nucleon (NN) correlations in this work) 
This problem is too complicated to be solved exactly in a consistent quantum-mechanical framework.
As a result, FSI is often one of the principal sources of uncertainty in neutrino oscillation results~\cite{Acero:2019ksn,Abe:2018wpn}.  A class of models called intranuclear cascade (INC) have been developed over many years. Foundations for the semi-classical treatment of FSI in terms of intranuclear cascade model were set by Serber~\cite{Serber:1947zza} and Metropolis~\cite{Metropolis:1958wvo, Metropolis:1958sb}. 
The propagation of hadrons is taken to be a series of independent binary collisions that are largely independent of each other.  Quantum corrections~\cite{Buss:2011mx,Cugnon:2016ghr,Ferrari:2005zk} have be introduced and detailed  agreement with a large variety of data has been obtained using a variety of approximations. 
Although theoretical arguments show that agreement of  cascade models with data is largely expected for situations when interaction probability is small, e.g. with nucleons of kinetic energy above 200~MeV~\cite{Yariv2007}, this agreement apparently extends to regions where the hadron-nucleon cross section is very large such as pion interactions when $\Delta$ resonance excitation is dominant~\cite{Ginocchio:1977pw}. Without an underlying quantum mechanical model, approximations are required.  In addition, experiments require codes which combine accuracy for most important quantities with numerical efficiency/speed. Even though GiBUU~\cite{Buss:2011mx} has an excellent dynamic model with the use of hadronic transport calculations, it is difficult to use in an experiment. 
Codes in use range from empirical~\cite{Dytman:2011zz} to theoretically motivated~\cite{Salcedo:1987md,Mancusi:2014eia} models. 

The basis of existing cascade codes is always free hadron-nucleon interactions with additional nuclear corrections and medium refinements.
The free hadron-nucleon interaction has been studied extensively and partial wave amplitude (PWA) fits to large data sets of cross sections and polarization quantities~\cite{said} are available.

One of the important choices required is the hadron-nucleus data set that is used to validate the models.  
Most often, cross sections with hadron beams on nuclear targets have become the sole FSI validation test and the total reaction cross section (\sigreac)~\cite{Carlson:1996ofz,PinzonGuerra:2018rju} is the most broad kind of data available.
Also known as the total inelastic or nonelastic cross section, it includes all final state channels except elastic and provides a single number for each energy that measures the total strength for important interactions.
Use of \sigreac for validation is straightforward because a large body of data is available and each generator has a subsidiary code using the same FSI code as in the neutrino-nucleus simulation to reproduce these data~\cite{Dytman:2011zz,Golan:2012wx,PinzonGuerra:2018rju}.  
The alternative is to use transparency data where the hadron is ejected from a nucleus by non-interacting probes such as an electron~\cite{Garino:1992ca,Pandharipande:1992zz} or a neutrino~\cite{Niewczas:2019fro}.
The advantage of this choice is that the final state hadron is produced in the same way as in neutrino oscillation measurements. 
Experimentally, the difficulties come from tagging the FSI hadron in an objective way.
Specifically, transparency is a ratio of hadrons in the final state with and without FSI.
Measurements to date use electron beams and  either tag the struck proton~\cite{Garino:1992ca} using the quasielastic interaction or normalizing the experimental result with a separate calculation of the result with no FSI~\cite{Dutta:2012ii}.
The latter choice is often used for high energy hadrons in studies of color transparency~\cite{Dutta:2012ii} where the goal is to find differences between data and nominal calculations as a function of $Q^2$ as evidence for unusual medium effects.
This article is aimed at issues important for neutrino oscillations~\cite{Acero:2019ksn,Abi:2020evt} and therefore focuses on particles of kinetic energy 1 GeV and lower.  There is also a large body of precise hadron-nucleus differential cross section data.  Comparison with these data will be a second step in benchmarking MC FSI models.  

The relationship between total reaction cross section and transparency is interesting, but not often studied.  
Pandharipande and Pieper~\cite{Pandharipande:1992zz} examined the theory input needed to describe proton transparency in the quasielastic region~\cite{Garino:1992ca}.
They note the importance of Pauli blocking, medium effects, and short-range nucleon-nucleon correlations.  
Ref.~\cite{Niewczas:2019fro} was devoted to studying the ability of the NuWro neutrino event generator~\cite{Golan:2012wx} to describe existing proton transparency data.
They also note the importance of medium corrections and short-range nucleon-nucleon correlations.
The GiBUU neutrino-nucleus model~\cite{Leitner:2009ke} was compared to proton transparency data with an emphasis on high momentum transfer, $Q^2>$ 1~$(GeV/c)^2$. 
Isaacson, et al.~\cite{Isaacson:2020wlx} recently introduced a novel FSI model based on ideas proposed in Ref.~\cite{Benhar:2003ka} and quantum Monte Carlo (QMC) calculations by generating a sample of realistic nuclear configurations within this framework.  They examined both $\sigma_{reac}$ and transparency with the goal to access the role of nucleon-nucleon correlations. They report negligible impact of correlations on $\sigma_{reac}$ and a moderate effect on transparency.

The purpose of this article is to study the FSI effects of protons and $\pi^+$ in nuclei through a combined analysis of hadron beam interactions and neutrino hadron production transparency.
This study features results from three commonly used neutrino event generators - GENIE~\cite{Andreopoulos:2009rq}, NEUT~\cite{neut}, and NuWro~\cite{Golan:2012wx}.
Since authors of each code contribute to this work, detailed comparisons are possible.
Although comparisons among generators are becoming more common~\cite{tensions2016,Stowell:2016jfr,Le:2019jfy,Abratenko:2020sga,Abe:2020uub}, comparisons involving detailed understanding of the codes are not often available.
Each code is used to calculate $\sigma_{reac}$ and transparency.
Although $\sigma_{reac}$ has been measured for many targets~\cite{Carlson:1996ofz,PinzonGuerra:2018rju}, the focus here is on carbon and argon because of their common use in neutrino oscillation experiments.
This will give an indication of the dependence on choice of nucleus, usually called A-dependence.
Transparency measurements are much less common.
Some results for protons are published for carbon and heavier elements~\cite{Garino:1992ca,ONeill:1994znv,Dutta:2003yt,Rohe:2005vc,Dutta:2012ii}, but none for argon yet.
The goal of most previous measurements is the search for color transparency~\cite{Dutta:2012ii,Brodsky:1988xz}; therefore, most of the existing data is at high momentum and energy transfer to the residual nucleus.
For example, existing measurements for pion production~\cite{Qian:2009aa} are aimed in that direction and there is then no transparency data for pions at energies needed for the neutrino oscillations experiments.
A major goal of this study is to examine the theoretical effects that make $\sigma_{reac}$ and transparency differ and encourage more experiments. 
  Since no acceptance corrections are made to match the data, these will be referred to as {\it Monte Carlo} calculations of transparency.
  This choice was made to allow examination of the relationship between the two quantities independent of any experimental details.
We will argue that within the intranuclear cascade approach there is an inherent relation between total reaction cross section and transparency coming from the mean free path and nucleon density. However, additional theoretical effects can modify the relation and a combination of precise reaction cross section and transparency measurements can be used to investigate those effects.  We notice that the relation between transparency and \sigreac becomes less intuitive in the hadron propagation approach studied in Ref.~\cite{Isaacson:2020wlx}.  At low values of proton kinetic energy $\sigma_{reac}$ gets large contributions from "long-range interactions" as illustrated in Fig.~8 of \cite{Isaacson:2020wlx}, i.e. nucleons far away from nucleus center and can still have non-negligible probability to interact.

\section{Reaction cross section vs. transparency} 
\label{sec:xsec_vs_transparency}

\subsection{General considerations}
The relationship between $\sigma_{reac}$ and transparency is a long-standing issue in FSI calculations.  This is closely related with the question about the best way to describe protons propagating in nuclei~\cite{Garino:1992ca}.  The issue there was the conflict between optical-model potentials and nucleon-nucleon interactions with medium corrections and the goal of Ref.~\cite{Garino:1992ca} was to provide data to resolve this conflict. 
The work of Pandharipande and Pieper cited above~\cite{Pandharipande:1992zz} showed that the NN interaction method correctly reproduced proton transparency data~\cite{Garino:1992ca}.  Since the data for $\sigma_{reac}$~\cite{Carlson:1996ofz,PinzonGuerra:2018rju} is much more commonly available, studies to date for neutrino interaction event generators largely use that information.

Here, we will explore the model dependence of both reaction cross reaction and transparency.
Although both quantities are sensitive to the mean free path of the hadrons, transparency samples the nuclear medium more directly since they start propagating randomly throughout the nucleus.
On the other hand, many strongly interacting particles (esp. pions at energies where the $\Delta$ (P$_{33}$(1232)) resonance is important) only interact in the periphery of the nucleus and rarely sample the interior.
In reality, hadrons produced inside the nucleus are off-shell. Since all the codes used here other than GENIE INCL++ propagate hadrons on-shell with off-shell effects included only at interaction points, that aspect cannot be studied here completely.  INCL++ puts the propagating particle in a potential well (both Coulomb and nuclear) that depends on particle type and its energy.  Therefore, the momentum of the propagating particle is constant while its energy varies with position. 

It is interesting to note the intrinsic differences between the two quantities. 
A formation zone accounts for the possibility of a hadron propagating for some distance in something other than a normal state, e.g. a particle with small transverse dimension~\cite{Brodsky:1988xz}.
This will affect transparency but not $\sigma_{reac}$.
Medium effects are relevant because a bound nucleon (N) has different behavior than a free nucleon due to the surrounding particles.
Medium modifications are known to be important in the propagation of pions~\cite{Freedman:1982yp} and protons~\cite{Pandharipande:1992zz} where they are often handled within a polynomial expansion in nucleon density. 
They will affect the two quantities in somewhat different ways because they have a strong dependence on nuclear density, e.g. if $\sigma_{reac}$ is largely determined by hadrons interacting in the surface of the nucleus.
Nucleon-nucleon (NN) correlations are beyond what can be handled with typical local density approximation techniques  and are treated separately~\cite{Pandharipande:1992zz,Niewczas:2019fro}.
These have almost no effect on $\sigma_{reac}$, but can be significant for transparency.

Although the codes considered here are largely aimed at neutrino interactions, transparency can in principle be measured either with electron or neutrino beams.  
Since only a fraction of the codes have the capability to simulate electron scattering events, the simulations here use electron neutrino ($\nu_e$) neutral current (NC) interactions.  They have almost identical kinematic properties to electron scattering interactions, using the weak interaction instead of the electromagnetic interaction.
The main purpose is to start the propagation of hadrons at locations according the matter density under identical residual nucleus conditions.
For proton transparency, the neutral current elastic interaction is used as the principal interaction.  
For pion transparency, the neutral current resonance interaction is used.  


\subsection{Toy model}
\label{sec:toy}
In this section we introduce a mathematical model which relates the values of total reaction cross section with transparency. The model is valid for arbitrary projectiles, including  pions and nucleons. It contains all the features of the basic propagation model, neglecting nucleon-nucleon correlations and formation zone effects which have impact on the measured transparency and can therefore be potentially considered as a tool to quantify those effects.  We also neglected the effects coming from Pauli blocking, or more generally from the dependence of the cross section on local density.

In the toy model, a nucleus is fully characterized by a density profile $\rho(r)$ assumed to be radially symmetric. It  satisfies the normalization condition

\begin{equation}
    \int \rho(r) d^3r = 4\pi\int r^2\rho(r) dr=A
\end{equation}
where $A$ is the atomic number.
    We assume that the bulk of the nuclear density is closed within a sphere of radius $R$. The value of $R$ is chosen so that its increase has no significant impact on the calculated values of transparency and \sigreac\ (4.65~fm for carbon and 7.1~fm for argon).

\subsubsection{Reaction cross section}
\label{sec:toy:reaction}

We assume a uniform flux of projectiles hitting a nucleus from the outside of the sphere of radius $R$. The goal is to calculate the probability $P_{reac}$ that an interaction happens during passage through the nucleus for a given value of the impact parameter. A product of the average value of $\langle P_{reac}\rangle$ with the geometric cross section $\pi R^2$ defines the reaction cross section. 

Technically, it is easier to calculate $1-\langle P_{reac}\rangle$, the average probability that the projectile travels through nucleus without interaction. Locally, the probability to move over the distance $dz$ without interaction is 
${\rm exp}{\{-\rho\sigma dz\}}$.
In the toy model we disregard the difference between proton and neutron local densities ($\rho$) and local cross sections ($\sigma$).

\begin{figure*}[th!]
    \centering
     \includegraphics[width=0.48\textwidth]{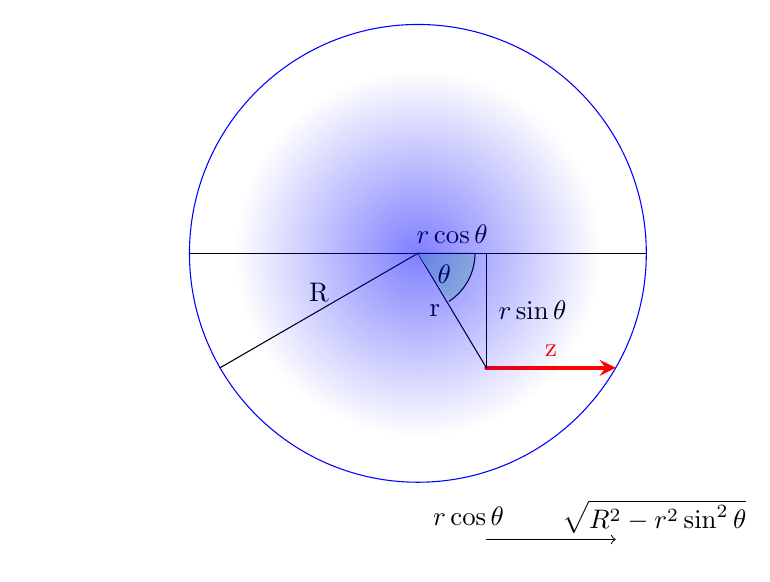}
     \includegraphics[width=0.48\textwidth]{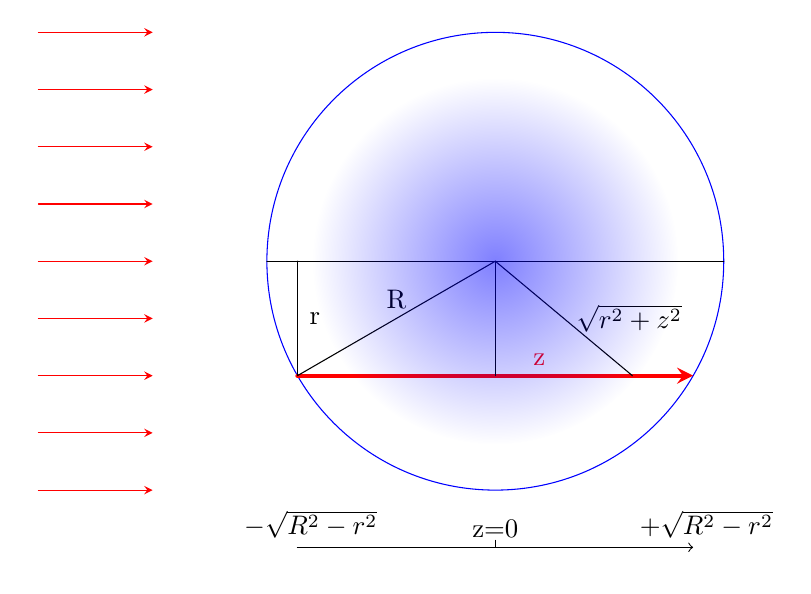}
    \caption{Computation of transparency (left) and reaction cross section (right) in the toy model.}
    \label{fig:toy-computations}
\end{figure*}

Taking into account that for each projectile the distance travelled inside nucleus is determined by the impact parameter $r$, we get the following expression (see the right side of Fig.~\ref{fig:toy-computations}):

\begin{eqnarray}
\sigma_{reac} & = & \pi R^2 - 2\pi \int_0^R dr\ r \nonumber \\&\cdot &{\rm \exp}\left\{ -\int_{-\sqrt{R^2-r^2}}^{\sqrt{R^2-r^2}}  \sigma\rho \left(\sqrt{z^2+r^2}\right) dz\right\}
    \label{toy-reaction}
\end{eqnarray}


\subsubsection{Transparency}
\label{sec:toy:transparency}

In this computation a trajectory starts at a point inside nucleus selected at random with the probability density given by $\rho (\vec r)$.

The computations lead to the following result for transparency ($T$) (see the left side of  Fig.~\ref{fig:toy-computations}) \cite{Pandharipande:1992zz}:

\begin{eqnarray}
    &T&=\frac{2\pi}{A}\int_{-1}^{+1}d(\cos\theta) \int_0^R dr\ r^2 \rho(r) 
    \nonumber \\ 
    & \cdot & {\rm exp}\left\{-\int_{r\cos\theta}^{\sqrt{R^2-r^2\sin^2\theta}}dz \sigma \tilde\rho \left(\sqrt{z^2+r^2\sin^2\theta}\right) \right\}
    \label{toy-transparency}
\end{eqnarray}


In the above formula the tilde in $\tilde\rho$ accounts for the fact that the numbers of spectator nucleons are different in nucleon reaction cross section and transparency computations. In the case of reaction cross section it is $A$ while in the transparency it is $A-1$. Thus $\tilde\rho = \frac{A-1}{A}\rho$. 


\subsubsection{Reaction cross section to transparency ratio}

\begin{figure}[th!]
    \centering
    \includegraphics[width=0.48\textwidth]{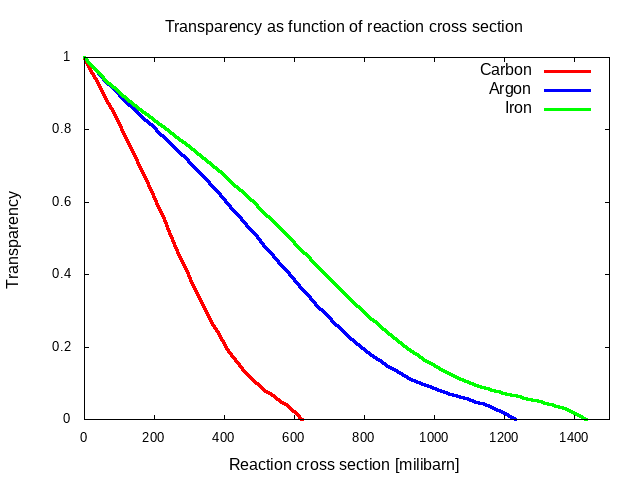}
    \caption{Transparency as function of reaction cross section in the toy model}
    \label{fig:r2t_toy}
\end{figure}

Absolute values of reaction cross section and transparency depend on the microscopic cross section $\sigma$ entering Eqs.~\ref{toy-reaction} and \ref{toy-transparency}. The exact value of $\sigma$ is not known but we can use information from Sects.~\ref{sec:toy:reaction} and  \ref{sec:toy:transparency} to eliminate $\sigma_{reac}$ and determine a function $T(\sigma_{reac})$. This was done for three realistic density profiles - carbon, argon and iron~\cite{DeJager:1974liz}. Results are shown in Fig.~\ref{fig:r2t_toy}. In the limit of reaction cross section going to zero, the transparency approaches the value one. In the other extreme case, when the reaction cross section approaches the maximal possible value of the geometric cross section ($\pi R^2$), the transparency goes to zero. In the intermediate region transparency is a monotone function of reaction cross section. Since this statement does not depend on any assumption about projectile kinetic energy, one can expect that if the projectile kinetic energy dependence of the reaction cross section shows a local minimum, the transparency should exhibit a local maximum and vice versa.

\subsubsection{Comparison with GENIE results}
\begin{figure}[th!]
    \centering
    \includegraphics[width=0.48\textwidth]{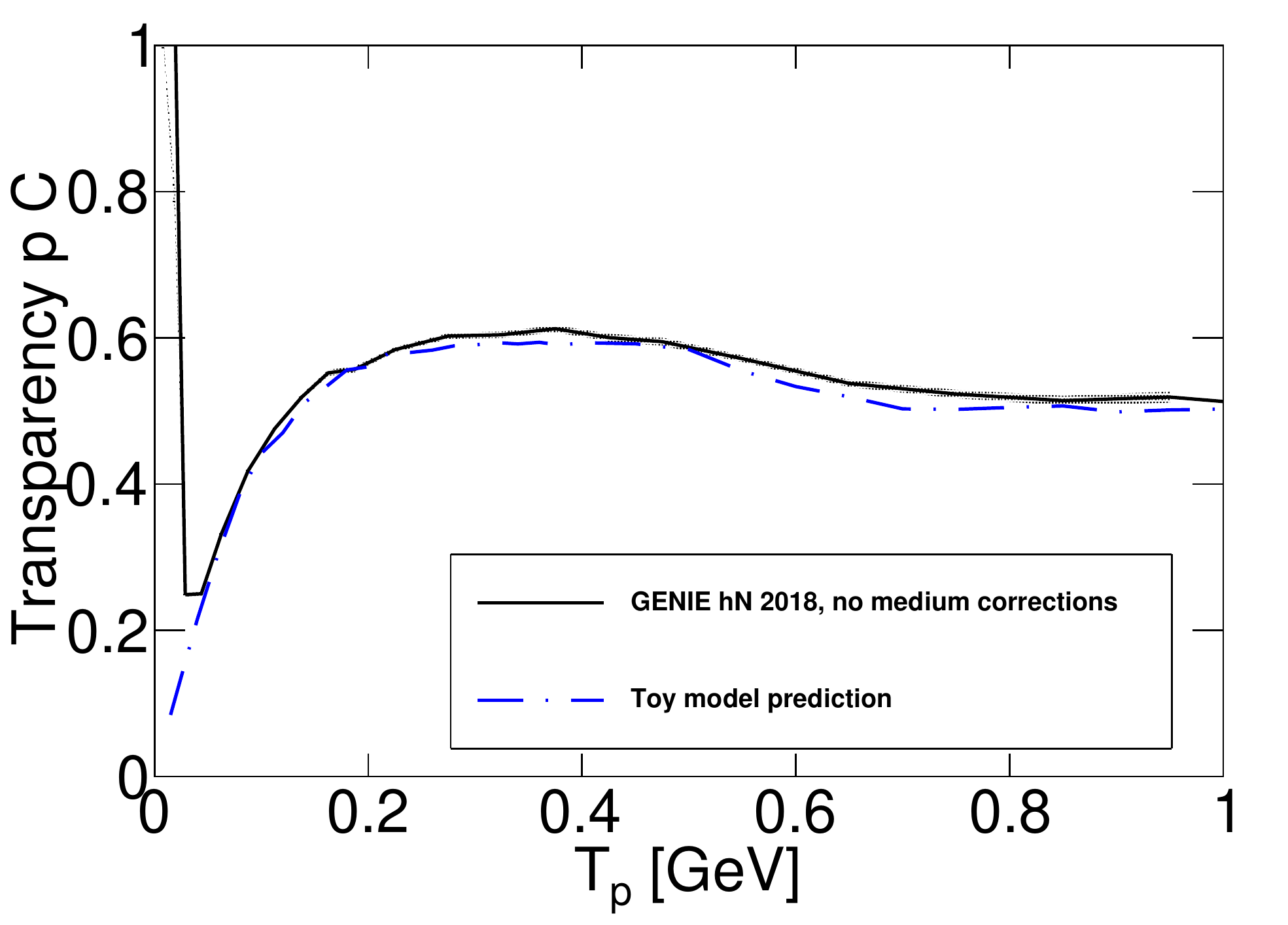}
    \includegraphics[width=0.48\textwidth]{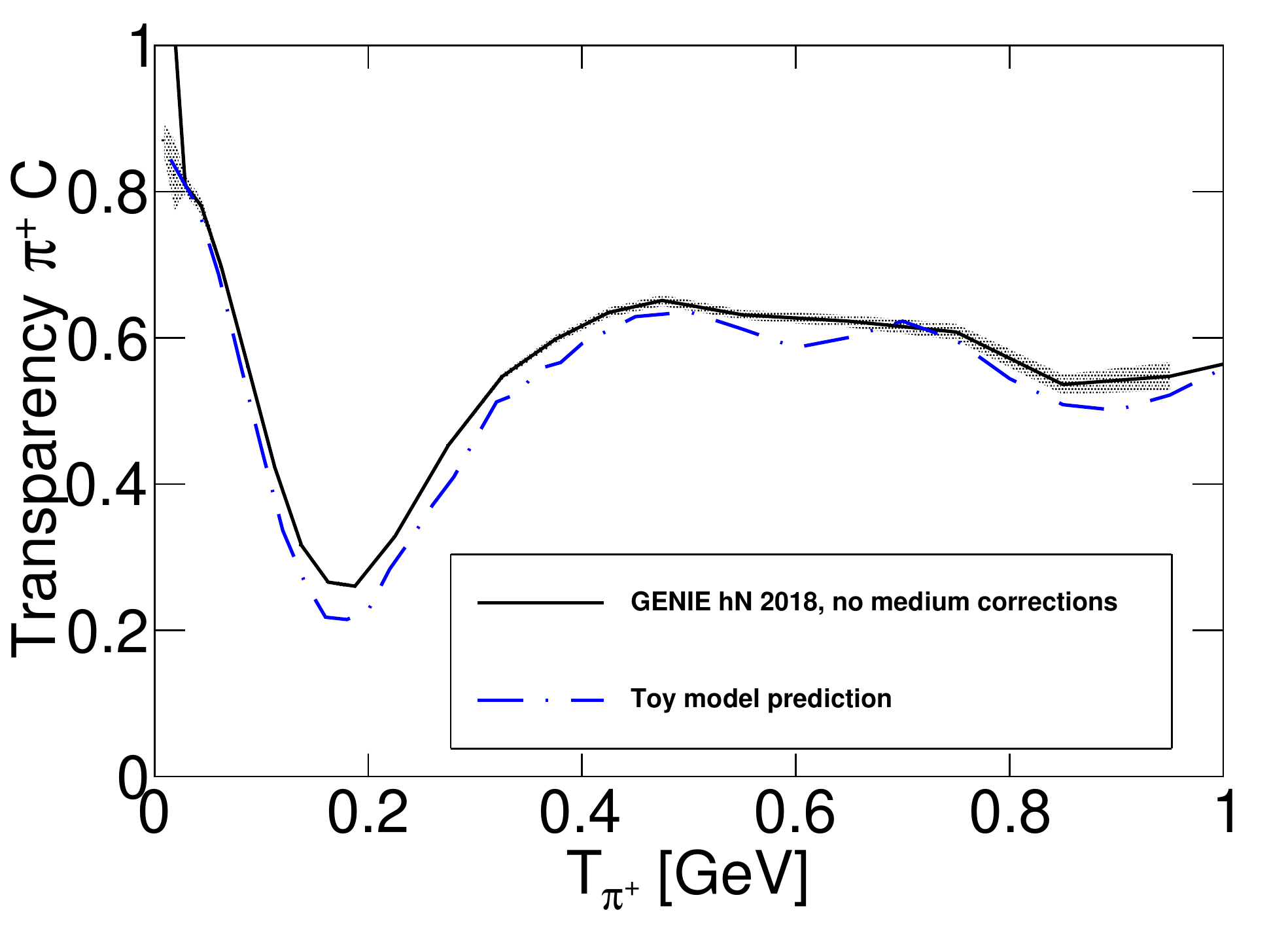}
    \caption{Transparency for proton and $\pi^+$ for carbon with results using bare GENIE hN2018 FSI model and the toy model using GENIE reaction cross section and the transparency/\sigreac ratio as explained in the text. In the GENIE simulation, all medium corrections and formation zone effects have been removed. The statistical error associated with GENIE predictions is represented with a grey band.}
    \label{fig:pC-piC-toy-genie}
\end{figure}

The toy model presented above results in a simple analytical equation which incorporates the most basic physics input.
More complicated codes can reproduce these results with suitable simplifications.
This can be studied in GENIE because of its modular design.  
 The toy model curve in Fig.~\ref{fig:pC-piC-toy-genie} was obtained by transforming GENIE reaction cross section results into transparency with the toy model ratio in Fig.~\ref{fig:r2t_toy}. Fig.~\ref{fig:pC-piC-toy-genie} demonstrates that stripped-down GENIE agrees well with the toy model for proton and $\pi^+$ transparency results in carbon. The GENIE transparency result goes to 1 below 20 MeV because a cutoff that was introduced (see Sect.~\ref{sect:GENIE} for details). The toy model and GENIE results use slightly different charge distributions, both consistent with electron scattering data~\cite{DeJager:1974liz}.  A general conclusion about the toy model is that it can be a useful tool to investigate nuclear effects modifying transparency but not reaction cross section. We adopt the approach where the reaction cross section is used as input to obtain results shown in Fig.~\ref{fig:r2t_toy}.

\section{Models} 
\label{sec:models}
 
 Event generator codes produce events from interactions of neutrinos, electrons, and hadrons with nuclei using similar core methods.  
 For this article, the focus is on propagation of protons and $\pi^+$ in nuclei.  
 GENIE hA2018 and hN2018, NuWro, and NEUT all treat the nucleus as an ensemble of largely independent particles.  These nucleons have momenta which typically come from a local Fermi gas distribution.  That means the momentum distribution is uniform in momentum space up to a maximum value (the Fermi momentum) which depends on position inside the nucleus:
\begin{equation}
p_F(\bf{r}\rm)=\left[\frac{3}{2}\pi^2\rho(\bf{r}\rm)\right]^{\frac{1}{3}},
\label{eq:fermimom}
\end{equation}
 Thus, the high momentum nucleons tend to be in the center of the nucleus.  This is a significant improvement beyond global Fermi gas models where the Fermi momentum is independent of position.
 Nucleons are in a bound system; this can be accounted for by giving the struck nucleon a shift in mass leading to an effective mass which is smaller.  Binding energy values are historically set by fits to high energy electron quasi-elastic scattering scattered electron energy distributions~\cite{Moniz:1971mt} where the correction is the shift of the peak position from the value appropriate for a free nucleon.
 INCL++~\cite{Mancusi:2014eia,Boudard:2012wc} has a more sophisticated nuclear model~\cite{Cugnon:2016ghr} where all nucleons are placed in a square well potential whose depth and range depend strongly on the nucleon's position and momentum. There are no collisions and the nucleons are constrained to simultaneously have a momentum distribution similar to the local Fermi gas~\cite{Mancusi:2014eia} and a spatial distribution according standard values~\cite{DeJager:1974liz}.  Tuning to a variety of hadron-nucleus scattering data is also important~\cite{Mancusi:2014eia}.  The result is an improved momentum distribution and binding energy correction.  Although the binding energy correction is clearly improved over any Fermi gas value by having a natural dependence on position and momentum, the result of their nuclear model has the highest momentum nucleons at the surface~\cite{Mancusi:2014eia} rather than in the interior.
 
 GENIE hA2018 and hN2018, NuWro, and NEUT all use a semi-classical approach where the hadron is assumed to be free and move in straight lines as it propagates. 
 The governing equation then comes from an evaluation of mean free path $\tilde\lambda$ which depends on position within the nucleus and the total cross section for interacting with nucleons.
 Hadrons propagating through a nucleus are moved in steps in position. The probability of traveling a distance $\lambda$ (small enough to satisfy an assumption of constant nuclear density $\rho$) without interaction is:

\begin{equation}
 P(\lambda) = e^{-\lambda / \tilde\lambda}
 \label{eq: fsip}
\end{equation}
where 
\begin{equation}
 \tilde\lambda = \left(\sigma_p \rho_p(r) + \sigma_n \rho_n(r)\right)^{-1}\label{eq:mfp}
\end{equation}
and $\sigma_{p}$ ($\sigma_{n}$) is the total cross section for interaction on either a proton ($p$) or neutron ($n$). Pauli blocking can either be included in average by reducing values of $\sigma_{p/n}$ or can be checked for each microscopic interaction. To satisfy the assumption of (approximately) constant density, the particles are moved by a small step with typical values from 0.05~fm (GENIE) to $0.2$~fm (NEUT and NuWro). Tests show that a range of step values give no visible impact on the final results but smaller steps make the computations slower. Using the probability calculated in Eq.~\ref{eq: fsip} at each step it is decided if an interaction is going to happen.  Although the GENIE INCL++ model is different in a variety of ways from the other approaches, it is still based on the mean free path concept for the energies considered here.
For INCL++, the entire hadron-residual system changes through time steps and interactions occur when the distance of separation is less than $\sqrt{\sigma_{hN}/\pi}$ fm~\cite{Mancusi:2014eia}.  A different method of propagating hadrons is proposed in Ref.~\cite{Isaacson:2020wlx}.  A configuration of nucleon positions is generated with the QMC method where all nucleons are interacting with realistic potentials. Hadrons to be tested (protons in their work) propagate as point-like on-shell particles in small time steps. Interactions can occur if there are target nucleons inside a cylinder with symmetry axis along the propagating nucleon momentum vector.  The cylinder's transverse size is the same as the INCL++ sphere of closest approach. In order for the interaction to be generated, hit nucleon initial momenta are sampled from local/global Fermi gas model density information and final momenta are checked for the Pauli blocking.  This method of propagating nucleons and generating interactions becomes very different from what is presented here at low values of nucleon kinetic energy when the nucleon-nucleon cross section is large.
 
For transparency computations in this work the hadron starting points are generated according to a density distribution as determined from elastic electron scattering.
To make sure that all propagating particles are deposited in the nucleus the same way, identical interactions are chosen for all the generators -- neutral current elastic (NCEL) interaction to produce a propagating proton and the neutral current resonance (NCRES) or deep inelastic scattering (NCDIS) interaction for $\pi^+$.
 The choice could be either electron or neutrino beams because each has essentially no interaction with the nucleus until the principal interaction.
 In this case, we chose neutrino interactions because not all participating codes have working electron capabilities.
 We also choose neutral current (NC) $\nu_e$ interactions to avoid issues with an improperly prepared residual nucleus.
 For example, a $\nu_\mu$ charged current (CC) interaction would have both an improper residual nucleus because of the conversion of a neutron to a proton and an energy imbalance due to the emission of a muon which has significant mass.
It is important to note that the transparency simulations here do not reproduce the requirements of the measurements~\cite{Garino:1992ca,ONeill:1994znv,Dutta:2003yt}.  Therefore, these results should have a qualifying label such as  {\it Monte Carlo transparency} and will only be indirectly compared with data in Sect.~\ref{sec:comparisons}.
 
 Nuclear effects can be included in a semiclassical model in a number of ways. 
 Various medium effects are added in accord with external models and data.
 They typically use a local density approximation.
 
 There are two interesting effects that modify theoretical results for transparency but not for reaction cross section. The first is {\it formation zone/time.}
  Formation time is also included in Ref.~\cite{Isaacson:2020wlx}.  The second effect is nucleon-nucleon correlations. They are studied in detail in Ref.~\cite{Isaacson:2020wlx} and  implemented in an approximate way in NuWro.

To summarize, nuclear effects are all handled in similar ways by the various codes.
However, different implementation choices have been made.
They are summarized in Table~\ref{tb:elasmodels} for protons and Table~\ref{tb:resmodels} for pions. These effects vary in size among the codes and will be studied in detail in Sec.~\ref{sec:modeldep}.

 \begin{table} [h]
\centering
\caption{Brief summary of model components for protons in NCEL  interactions that influence \sigreac and transparency in different ways. }
{\renewcommand{\arraystretch}{1.2}
\begin{tabular} {ccccc}
\hline\hline\noalign{\smallskip}
Generator & Pauli  & medium  & NN  & formation \\
& blocking & effects & correlations & zone \\
\hline\hline
GENIE hA2018 & none & Ref.~\cite{Pandharipande:1992zz} & none & none\\
GENIE hN2018 & none & Ref.~\cite{Pandharipande:1992zz} & none & none \\
GENIE INCL++ & yes & Ref.~\cite{Boudard:2012wc} & none & none \\
NuWro 19.02  & yes & Ref.~\cite{Pandharipande:1992zz} & Ref.~\cite{Pandharipande:1992zz} & none\\\
NEUT v5.4.0.1 & yes & none & none & none\\
\hline\hline
\end{tabular}}
\label{tb:elasmodels}
\end{table}

 \begin{table} [h]
\centering
\caption{Brief summary of model components for pions in NCRES interactions that influence \sigreac and transparency in different ways. }
{\renewcommand{\arraystretch}{1.2}
\begin{tabular} {cccc}
\hline\hline
Generator & Pauli  & medium  & formation \\
& blocking & effects & zone \\
\hline\hline
GENIE hA2018 & none & none & none\\
GENIE hN2018 & none & Ref.~\cite{Salcedo:1987md} & none \\
GENIE INCL++ & yes & Ref.~\cite{Boudard:2012wc} & none \\
NuWro 19.02  & yes & Ref.~\cite{Salcedo:1987md}  & none\\
NEUT v5.4.0.1 & yes & Ref.~\cite{Salcedo:1987md} & Ref.~\cite{Ammosov:2001} \\
\hline\hline
\end{tabular}}
\label{tb:resmodels}
\end{table}

\subsection{NEUT}
\label{sec:neut}
Like the other generators, pion and nucleon FSI are simulated using a custom semiclassical INC model. The nuclear density function is of Woods-Saxon type. The mean free paths for
low momentum pion interactions in the Delta region are calculated using the
prescriptions by Salcedo et al.~\cite{Salcedo:1987md}. 
These authors proposed a model of $\Delta$ propagation in a finite nucleus.  The model includes a modification of $\Delta$ self-energy due to medium effects via the local density approximation. 
This was an appropriate way to reproduce the pion-nucleus data available at the time, and was aimed for Monte Carlo simulation.
The main advantage is an accurate description of pion cross sections for a wide range of nuclei.
This approach was first adopted by NEUT and later included in NuWro and GENIE. 
The mean free paths for pions with momentum above 500 MeV/c), are obtained using the free  pion-nucleon scattering data. 
In the calculation of the mean free paths for the low momentum pions, the Local Fermi
Gas model is used and thus Fermi motion of the nucleon in the nucleus was taken into account.
Pauli blocking was implemented by requiring 
the nucleon momentum after the interaction to be larger than the local Fermi surface momentum ($p_F(r)$) (see Eq.~\ref{eq:fermimom}).

In order to improve the match of the simulation to pion-nucleus total cross section data, 
energy independent normalization factors were introduced for each pion-nucleon interaction type, 
e.g. charge exchange or absorption.  These factors were then fit to various 
pion-nucleus scattering data sets for pion energies up to 2  GeV.~\cite{PinzonGuerra:2018rju} 

In the determination of the kinematics of the scattered pions, the results
 of phase shift analysis obtained from $\pi-N$ scattering experiments~\cite{Rowe:1978fb}
are used.  Although the mean free paths are determined according to Salcedo-Oset, the outgoing kinematics were modified according to medium corrections as suggested by Seki et al.\cite{Seki:1983sh}.
Also, Pauli blocking effect is taken into account to be consistent with the mean free path.

\begin{figure}[th!]
    \centering
    \includegraphics[width=0.48\textwidth]{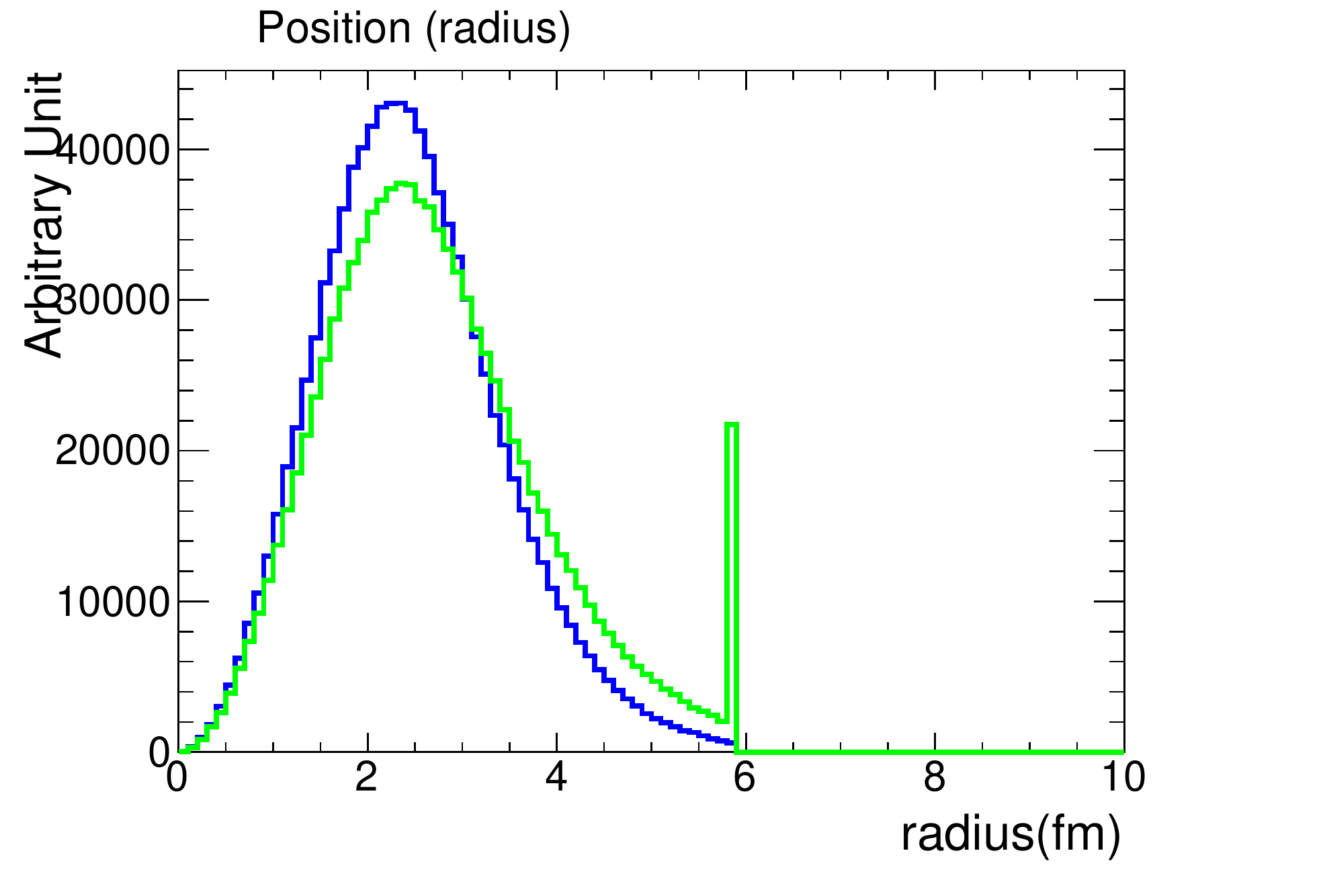}
    \caption{Effects of the formation zone in NEUT. The plot shows the interaction position (distance from the nucleus center) distributions of the neutrino (blue) and the starting position of the pion (green) coming from the neutrino interaction. The peak at 6 fm is at the outer edge of the Carbon nucleus in the simulation. Because of the formation zone effect, the production position of pion is shifted to the region of lower density. }
    \label{fig:piC-neut-position-formzone}
\end{figure}

 FSI for nucleons uses the differential cross-sections of nucleon scattering in nucleus from the work by Bertini et al.~\cite{Bertini:1972vz}, which was based on various experimental results.  Elastic scattering and one or two $\Delta(1232)$  production processes are included. 
In order to simulate the $\Delta(1232)$ production, a simple isobar production model~\cite{Lindenbaum:1957ec}
is used. The produced pions are tracked using the pion re-scattering code from the point of generation. The treatments of the binding energy (effective nucleon mass) 
for nucleon scattering is quite different than either pion scattering
in NEUT or the other generators. In simulating the nucleon FSI in NEUT,
the effective nucleon mass ${M_N}^{eff}$ of the bound nucleon is set to be 
\begin{equation}
    {M_N}^{eff} = \sqrt{({M_N}^{free}-8 MeV/c^2)^2-(p_F^{surf})^2},
\end{equation}
where $\rm{M_N}^{free}$ is the on-shell nucleon mass and $p_F^{surf}$ is 
the local Fermi surface momentum (see Eq.~\ref{eq:fermimom}), respectively. While tracking the nucleon in the nucleus, 
an interaction with the other nucleon happens only when total energy is larger
than 2 $\times M_N^{free}$. This largely suppresses the low energy interactions of nucleons as shown in the later comparisons,
in Figs.~\ref{fig:reac-xs-trans-pC} and \ref{fig:reac-xs-trans-pC-zoom}.
In simulating hadron production processes such as single meson or multi-pion production and deep inelastic scattering, a formation zone effect based on SKAT data is included in NEUT~\cite{Baranov:1985mb,Ammosov:2001}. 
The production point of the hadrons for those interactions are shifted using the formation length ($L_{FZ}$), where
\begin{equation}
    L_{FZ}=p/\mu^2,
\end{equation}
p is the momentum of the hadron and $\mu=0.08$(GeV/c$^2$). The actual size of the shift is
determined as $L_{FZ}\times(-\log({\rm rand}[0,1])$, where rand[0,1] is a positive random number up to 1. Because of the geometrical effects,
this makes the distribution of the hadron production locations shift to the outer region of the nucleus as shown in Fig.~\ref{fig:piC-neut-position-formzone}. 
The density of the surrounding region is lower than the central part and thus, the interaction probability of those particles becomes smaller. 

\subsection{NuWro}
\label{sect:nuwro}

In NuWro the expected distance travelled by hadrons is calculated to be:

\begin{equation}
 \lambda = - \tilde\lambda\cdot\log(\mbox{rand}[0,1])
 \label{eq:distance}
\end{equation}
with $\tilde\lambda$ defined in Eq.~\ref{eq:mfp}. An interaction is generated at the distance $\lambda$ if its value calculated in Eq.~\ref{eq:distance} satisfies $\lambda < 0.2$~fm. 
If instead $\lambda \geq 0.2$~fm the particle is moved by a distance $0.2$~fm along its trajectory.
 
FSI effects are described by a custom made semiclassical intranuclear cascade (INC) model \cite{Golan:2012wx, Niewczas:2019fro}.
The essential ingredient of the model are hadron-nucleon microscopic cross sections.
For pions, in the kinetic energy range ${T_k = 85-350 \text{ MeV}}$, NuWro, like NEUT, uses the results of microscopic calculation from Ref.~\cite{Oset:1987re} implemented 
in the region of $T_k\leq 350$~MeV
as described in  Ref.~\cite{Salcedo:1987md}. The model parameters are tabulated as functions of nuclear density in intervals of $0.1\rho(r)/\rho_0$ and can be used for arbitrary nuclei once its density profile is known. $\rho(r)$, $\rho_0$ are local and saturation nuclear densities, respectively. The parameters are interpolated linearly to cover the whole range of nuclear density. Unlike in NEUT, no adjustment of the parameter values to the experimental data for pion absorption and reaction cross sections was done. 

Since the model in Ref.~\cite{Salcedo:1987md} is restricted to the energies up to the $\Delta(1232)$ region only, it is supplemented with a phenomenological model for 
kinetic energies above 350 MeV. Free pion-nucleon cross sections are taken from the available experimental data. 


For nucleons NuWro uses as a starting point a custom fit to the experimental free \mbox{nucleon-nucleon} cross sections, both elastic and inelastic.
The fraction of \mbox{single-pion} production within inelastic interactions was adjusted to follow the fits of Ref.~\cite{Bystricky:1987yq}. 

The in-medium modification of the elastic nucleon-nucleon cross sections is modeled using the results of the Pandharipande and Pieper study~\cite{Pandharipande:1992zz}, where two main effects come from Pauli blocking and \mbox{in-medium} nucleon effective mass.
The Pauli blocking is included on the \mbox{event-by-event} basis.

For inelastic \mbox{nucleon-nucleon} scattering NuWro adopts a phenomenological \mbox{in-medium} microscopic cross section ($\sigma^\ast_{\mathrm{NN}}$) parameterization~\cite{Klakow:1993dj}:
\begin{equation}
    \sigma^\ast_{\mathrm{NN}} = \left(1 - \eta \frac{\rho (r)}{\rho_0}\right) \sigma^{\mathrm{free}}_{\mathrm{NN}},
\end{equation}
where $\eta = 0.2$.

Following the guidance of Refs.~\cite{Pandharipande:1992zz,Benhar:2006hh,Cosyn:2013qe}, the NuWro cascade model includes {\sl correlation effects} coming from \mbox{nucleon-nucleon} correlations.
The main effect comes from \mbox{short-range} correlations.
The density that enters the mean free path formula is (for the details see Ref.~\cite{Niewczas:2019fro}): 

\begin{equation}
    \begin{split}
    \rho^{[1]}_{\mathrm{eff,IPSM}}(\vec{r}_2|\vec{r}_1)  =  \rho^{[1]}_\mathrm{A-1}(\vec{r}_2) g(|\vec{r}_{21}|) N(|\vec{r}_1|),
    \end{split}
    \label{eq:effdensity_sub}
\end{equation}
normalized to the number of spectator nucleons 

\begin{equation}
    \int
\mathrm{d}^3\vec{r}_2 \
\rho^{[1]}_{\mathrm{eff}}(\vec{r}_2|\vec{r}_1) = A-1.
\end{equation} $g(|\vec{r}_{21}|)$ is the \mbox{nucleus-dependent} pair distribution function~\cite{Pandharipande:1992zz} and $N(|\vec{r}_1|)$ is introduced to keep the global normalization condition. For the choice of $g(|\vec{r}_{21}|)$, we rely on distributions of \mbox{nucleon-nucleon} distances obtained in ab initio computations for light nuclei, including carbon~\cite{ANLdensity,Carlson:2014vla}.
For heavier nuclei including iron, we approximate $g(|\vec{r}_{21}|)$ by the \mbox{ab initio-calculated} infinite nuclear matter distributions $g_{\mathrm{inf}}(\rho_{\mathrm{avg}},|\vec{r}_{21}|)$ of Ref.~\cite{Pandharipande:1992zz}, evaluated at average nuclear density.
The model includes effects coming from different shapes of $g(|\vec{r}_{21}|)$ for nucleon pairs of the distinct isospin configurations.

For both NuWro and GENIE, formation time/zone effects are included in DIS events, therefore not included in the main results of this study.  A somewhat similar effect, originating from finite $\Delta$ life-time can also be applied in resonance events.  NuWro has this as an option that is not used in this study.  

\subsection{GENIE}
\label{sect:GENIE}
GENIE allows many configurations, each of which contains a different set of models.
In addition, many of the model sets have configurable parameters inside; all simulations here come from the version 3.0.6 public release with the exception of INCL++ which is in v3.2.0.
Each model set has a complete set of core interactions containing the most compatible models.
For example, the G18\_10a\_02\_11a model set has the local Fermi gas nucleon momentum distribution for all interaction models in it.
For this work a single GENIE model set was chosen to deposit a proton (NCEL) or $\pi^+$ (NCRES or NCDIS) within the nucleus as discussed above.
In version 3 of GENIE, there are multiple FSI models that can be used in conjunction with each set of primary interactions.
Three of them are used for the present study.
{\it hA} has been the traditional default model and {\it hN} was added in v3.\footnote{ The names are derived from the scope of each model.  {\it hN} is a traditional hadron-nucleon INC model and {\it hA} is a custom model which has a more empirical characterization of the effect of multiple hadron-nucleon interactions.} 
Both {\it hA2018} and {\it hN2018} have significant modifications which make them more complete than previous versions.
In particular, {\it hA2018} has deleted the hadron-nucleus elastic scattering process which has had the effect of significantly increasing the transparency while leaving \sigreac largely unchanged. 
Both use the same semi-classical stepping method according to mean free path as described above with a step size of 0.05 fm.
Neither use Pauli blocking to suppress interactions at lower energies. 
No formation zone effects are employed in any of these FSI codes.
All use a modified Gaussian spatial nucleon density function which is very similar to the Woods-Saxon function used by NuWro and NEUT.

The {\it hN} FSI model has very similar features to what is in NuWro and NEUT.
It uses the Salcedo-Oset~\cite{Salcedo:1987md} model for pions with kinetic energy below 350 MeV.
This model has nuclear medium modifications.
Above 350 MeV, unmodified free $\pi N$ interactions~\cite{said} are used.
Nucleon interactions are also based on nucleon-nucleon cross sections which are based on the phase shift fits of the GWU group~\cite{said} to a variety of data.
Nucleon FSI models use the Pandharipande-Pieper medium modifications~\cite{Pandharipande:1992zz} at all energies.
 The {\it hN} model uses an energy cutoff below which low energy hadrons don't propagate; the cutoff depends on the nucleus as $A^{0.2}$.

The {\it hA} model is more empirical.
The main differences with respect to {\it hN} come after a final state interaction is chosen.  The {\it hA} model has a single interaction which is heavily based on data instead of the multiple interactions of cascade models such as {\it hN}, NEUT, and NuWro employ.
It uses the same mean free path values as {\it hN} with the exception of the pion-nucleon cross section.
Although it uses the Pandharipande-Pieper model for nucleons, the pion interaction has no medium modifications.

The INCL++ model~\cite{Mancusi:2014eia,Boudard:2012wc} has been recently added.
Its development has been completely independent of the other codes discussed here and is being used in production codes for the first time.
The original goal of this model was a global capability to describe nucleon-nucleus interactions at energies spanning the range studied here.  Understanding the  spallation process was a major goal and it is one of the best existing codes~\cite{iaea}.
All particles are tracked as a function of time.
The interplay of radial and momentum dependence of the nucleon distribution is carefully considered~\cite{Cugnon:2016ghr}.  The propagating hadron and all nucleons in the residual nucleus are in a square well potential whose width depends on location.  The depth and radius were originally tuned to spallation data, but the radius was tuned to single nucleon emission data in a recent publication~\cite{Mancusi:2014eia}.  
By choosing constant well depth, the accounting for energy and momentum is much simpler.
The INCL++ nuclear model is therefore more sophisticated than the other codes used in this study and automatically has medium effects on both nucleons and pions.
An extra effect in INCL++ is the effective binding of the propagating hadron that decreases the kinetic energy.
Tracking of higher energy particles is similar to the other approaches here with a more sophisticated nuclear model.  After a stopping time (7.8 x 10$^{-22}$s for argon), propagation is handed over to a pre-equilibrium/compound nucleus model, ABLA07~\cite{Kelic:2009yg}.  Although this is important for low energy nucleons in the final state, it is unlikely to have a strong effect on the results in this work.
The $\Delta$ is a separate particle that propagates independently with competing interactions and decay possibilities.
The other codes don't have an explicit $\Delta$, but the propagating pion has effects of both the $\pi\rightarrow\Delta$ transition and medium effects according to Ref.~\cite{Salcedo-Oset}.

\section{Comparisons}
\label{sec:comparisons}
We now present a series of comparisons between total reaction cross sections and transparency for protons and $\pi^+$ interacting with carbon (specifically $^{12}C$) and argon ($^{40}Ar$) targets.  
The goal is to identify similarities and differences in results obtained with NEUT, NuWro and GENIE using codes that a user could employ.

\begin{figure}[th!]
    \centering
    \includegraphics[width=0.48\textwidth]{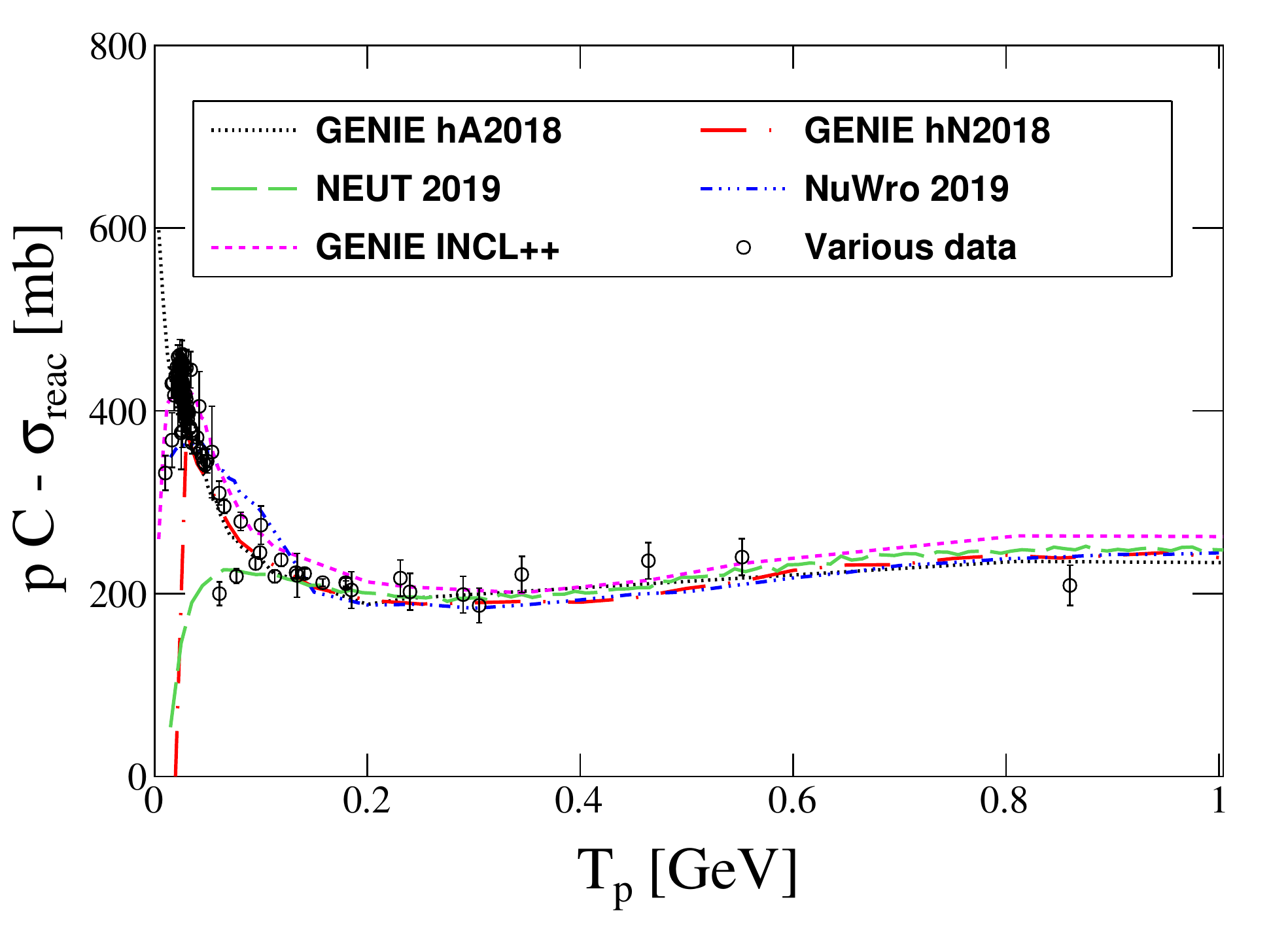}
    \includegraphics[width=0.48\textwidth]{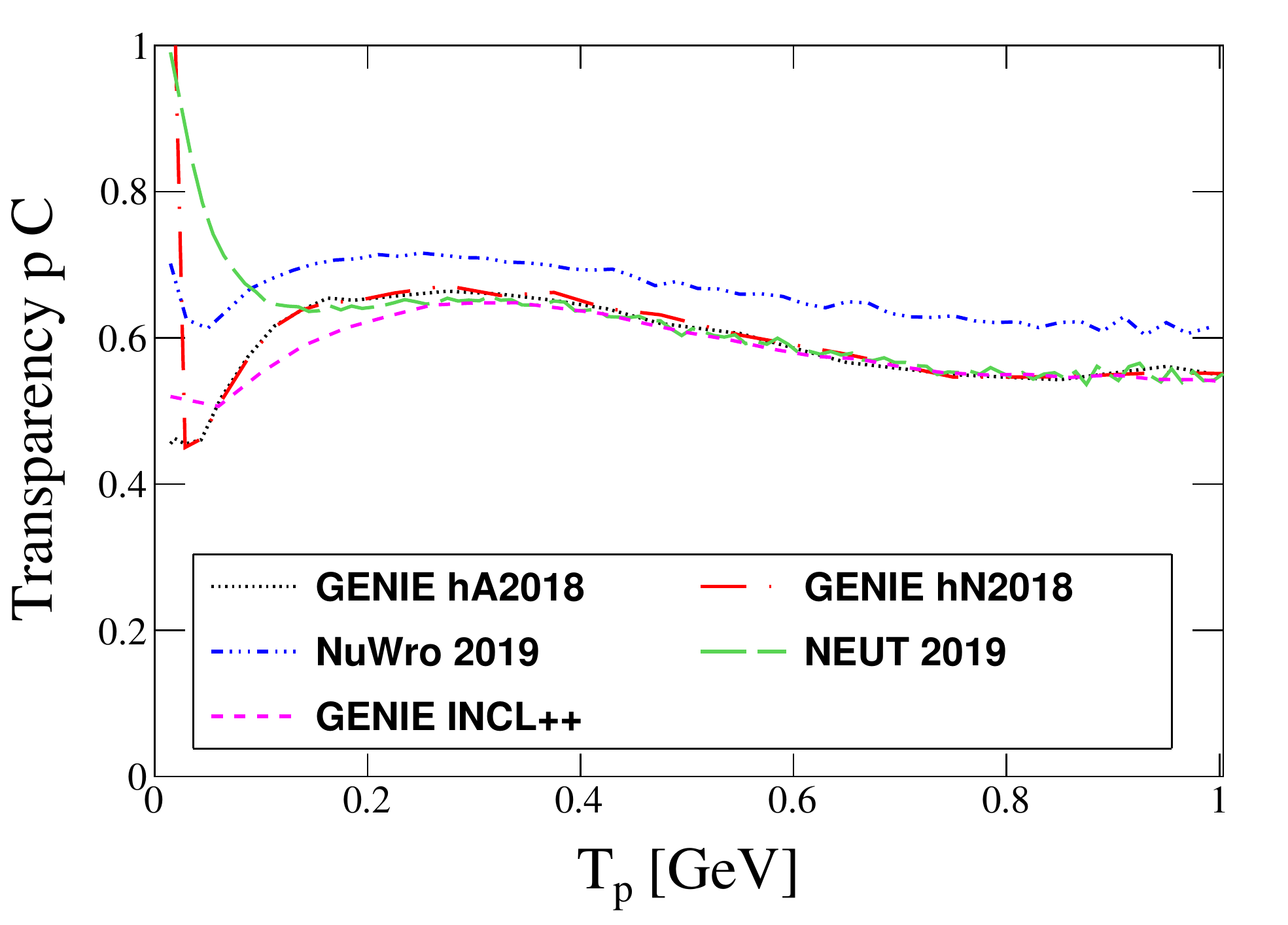}
    \caption{Total reaction cross section and transparency for proton-carbon.  Available data for \sigreac~\cite{Carlson:1996ofz} is shown along with calculations from GENIE, NuWro, and NEUT.  The transparency results are {\it Monte Carlo}, i.e. with no acceptance corrections.}
    \label{fig:reac-xs-trans-pC}
\end{figure}

Fig.~\ref{fig:reac-xs-trans-pC} 
provides a broad look at proton-carbon total reaction cross section and transparency up to 1 GeV kinetic energy.  Transparency data is not shown because the results are {\it Monte Carlo} values in that experimental acceptances are not taken into account (details are discussed later in this section).  No results of the calculations are shown below 20 MeV because most models are not aimed at energies that low.  It is notable that there is separation by proton energy in \sigreac, larger energies ($T_p  > \sim 200$ MeV) where the reaction cross section calculations tend to agree with each other and the data together with a lower energy region where data is less certain and the calculations diverge as well. 
For the higher energy region, the total cross section calculations are slowly rising and transparency calculations are slowly falling.
This relationship was first discussed in Sect.~\ref{sec:toy} and agrees with the predictions there.
Although the NuWro value for \sigreac 
agrees with the other calculations, the transparency prediction is above the others as NN short range correlations influence transparency but not reaction cross section (see Sect.~\ref{sec:modeldep}).

\begin{figure}[th!]
    \centering
    \includegraphics[width=0.48\textwidth]{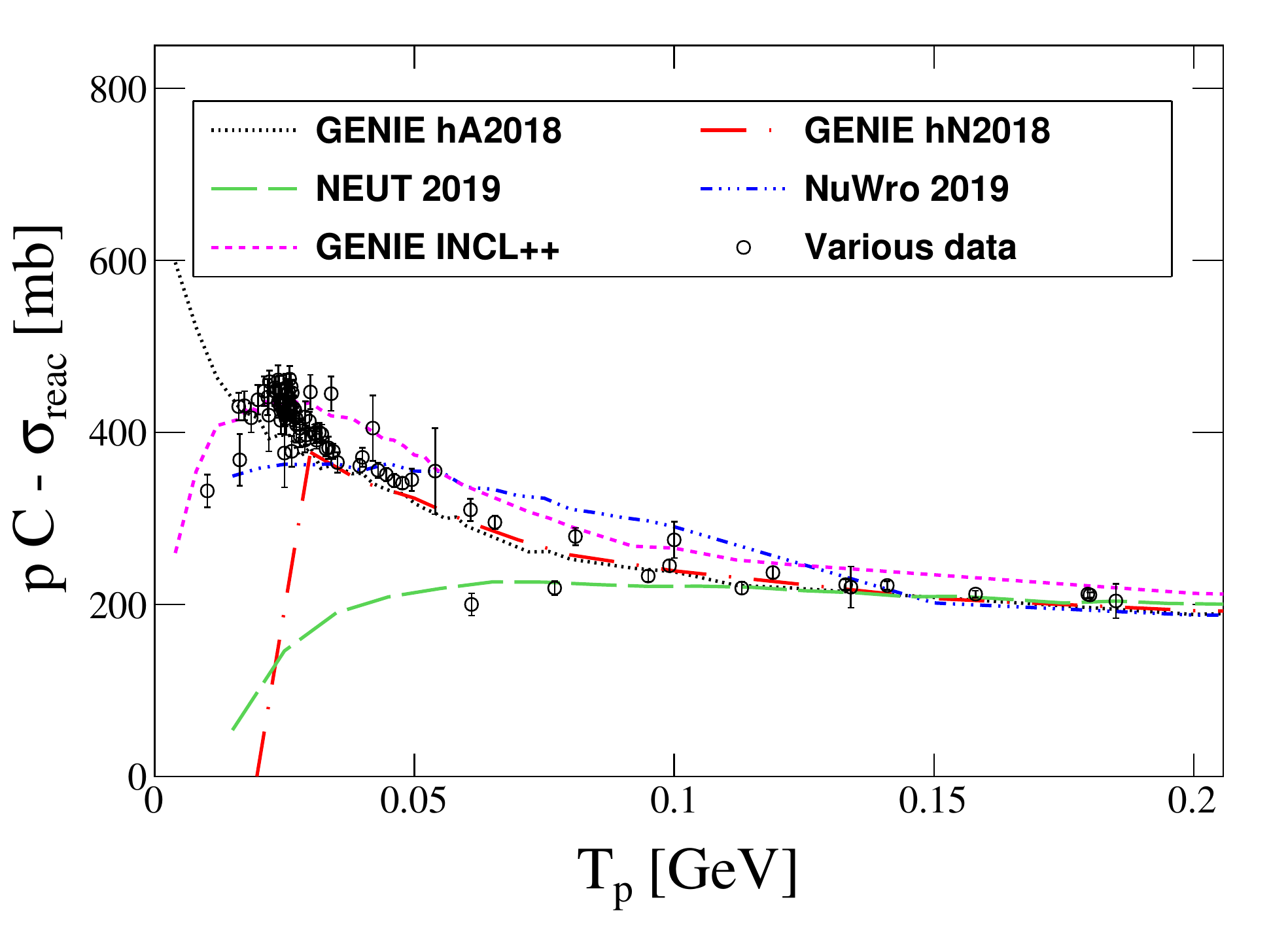}
    \includegraphics[width=0.48\textwidth]{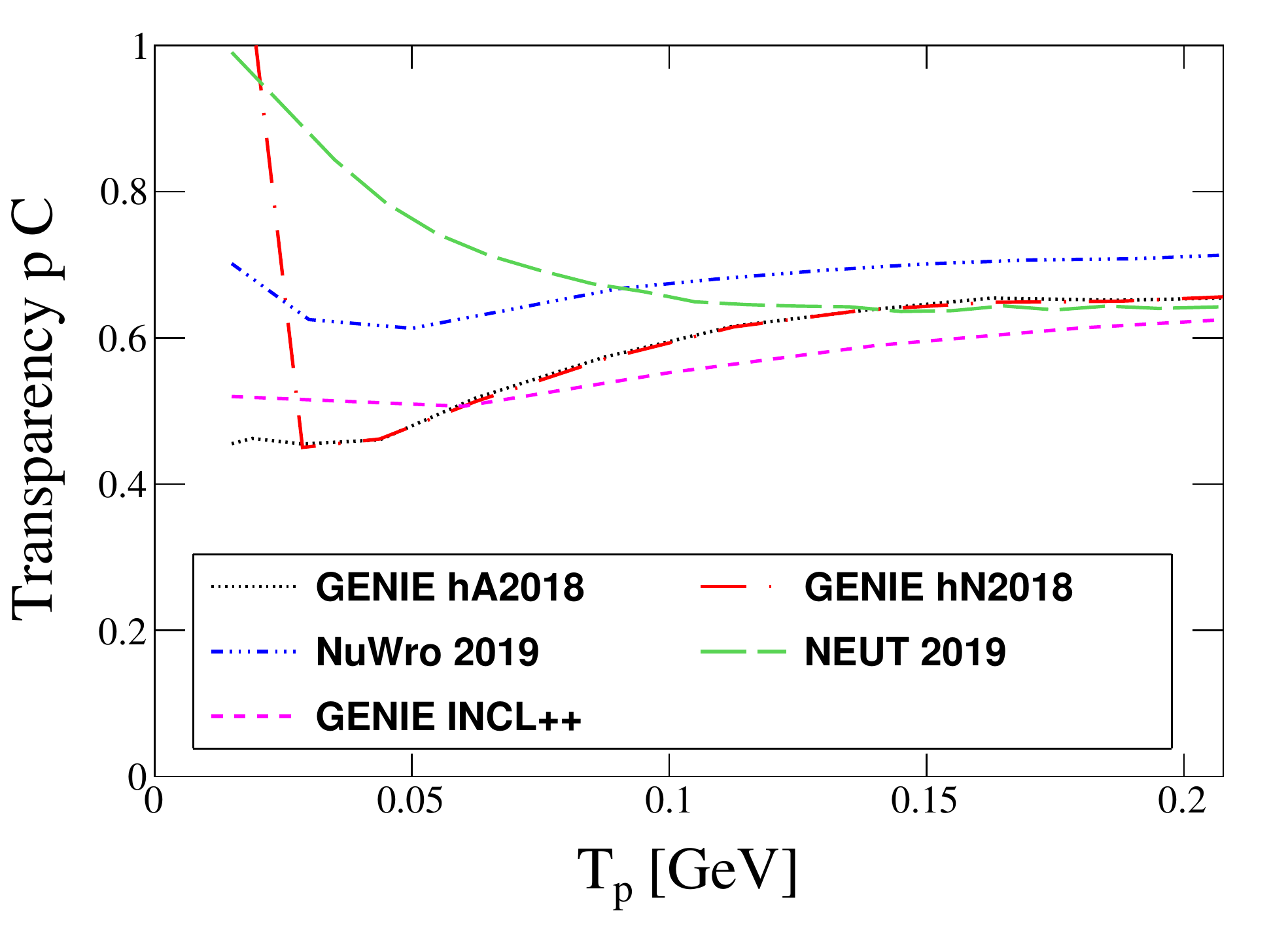}
    \caption{Total reaction cross section and transparency for proton-carbon, same as Fig.~\ref{fig:reac-xs-trans-pC} except for an expanded scale to show details.  Available data is shown along with calculations from GENIE, NuWro, and NEUT.}
    \label{fig:reac-xs-trans-pC-zoom}
\end{figure}

Fig.~\ref{fig:reac-xs-trans-pC-zoom} focuses on the lower proton energies so that details from Fig.~\ref{fig:reac-xs-trans-pC} become apparent.
The \sigreac data~\cite{Carlson:1996ofz} has some scatter because much of it comes from publications from 1960's and 1970's.  Overall, the data can be understood as a rise as the energy decreases to account for the increasing $pN$ cross section and the influence of compound nuclear processes.
The cross section peaks at $\sim$30 MeV and decreases at lower energies due to Pauli blocking, other in-medium effects, and Coulomb repulsion. 
The peaking is seen in NuWro and GENIE-INCL++ in good agreement with the data.  INCL++ has both compound nuclear processes and Coulomb effects, but both are absent in NuWro. 
Since both INCL++ and NuWro have Pauli blocking, this seems to be the most important contributor.
GENIE-hA2018 continues to increase to lower energies while GENIE-hN2018 follows the same trend but is cut off at 20 MeV.
Neither have Pauli blocking in the code versions used here and the {\it hN} code compensates for this with the empirical cutoff, as explained in Sec.~\ref{sect:GENIE}.
The NEUT simulation starts to fall off at about 80 MeV for both carbon and argon
because of the different treatment of binding energy and Pauli-blocking
as explained in the Sec. \ref{sec:neut}.
The corresponding transparency calculations for GENIE-hN2018 and NEUT have a rapid rise in transparency where \sigreac is rapidly decreasing as discussed in Sec.~\ref{sec:toy}.
The peak in \sigreac at about 40 MeV corresponds to a dip in transparency for NuWro and GENIE-INCL++.  

The detailed correspondences are also interesting.
GENIE-INCL++ has the most complete nuclear model and has the best agreement with the \sigreac data.
While GENIE-INCL++ is always below NuWro in transparency in Fig.~\ref{fig:reac-xs-trans-pC-zoom}, NuWro is sometimes larger and sometimes smaller for \sigreac.
Even though GENIE-hA2018 has a minimal set of nuclear corrections, it tends to agree with GENIE-INCL++ for energies larger than about 30 MeV.

\begin{figure}[th!]
    \centering
    \includegraphics[width=0.48\textwidth]{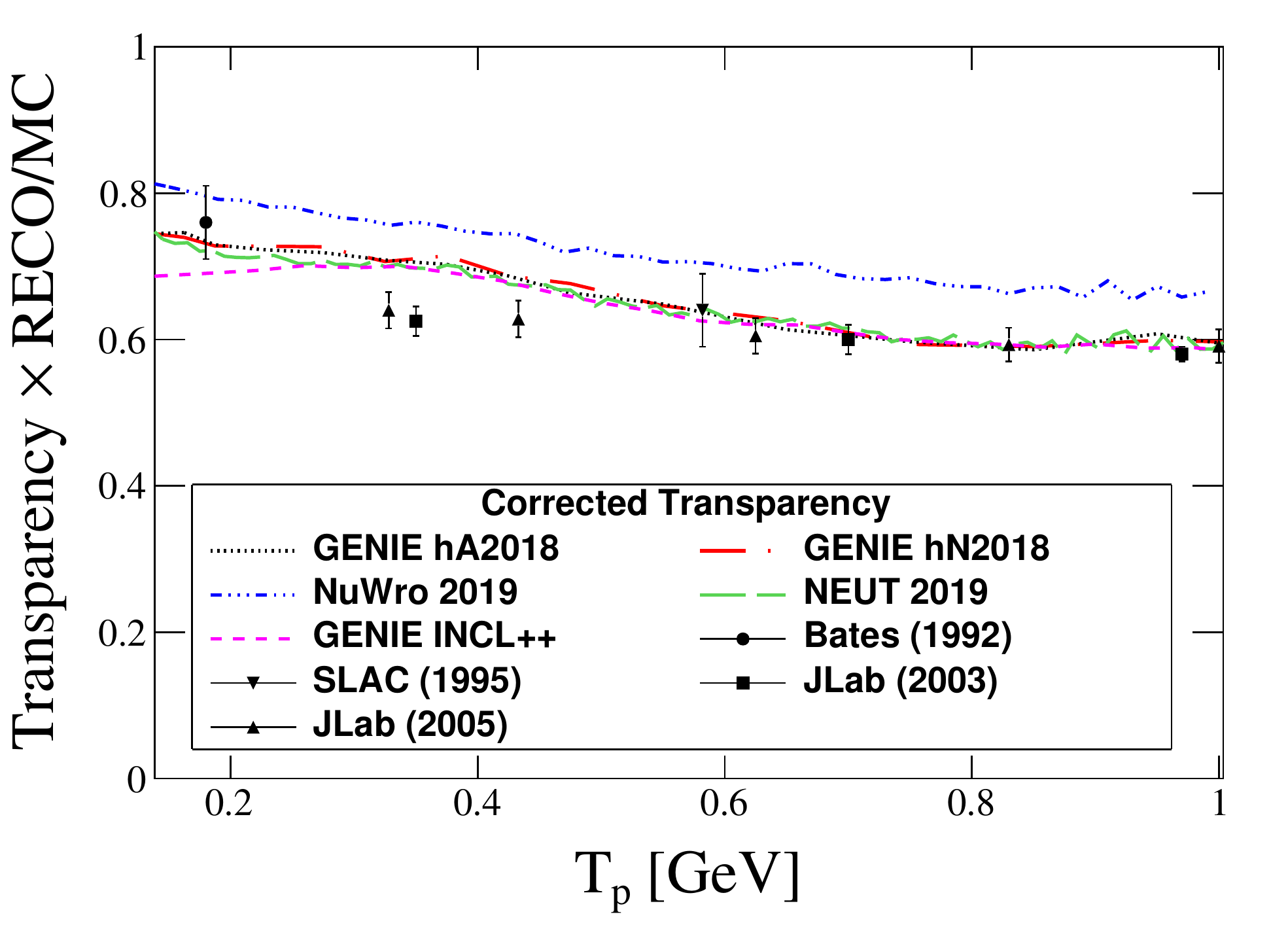}
    \caption{Transparency for proton-carbon where the calculations have been corrected according to acceptance effects as determined in Ref.~\cite{Niewczas:2019fro}.  Available data~\cite{Garino:1992ca,Dutta:2003yt,ONeill:1994znv,Rohe:2005vc}, is shown along with calculations from GENIE, NuWro, and NEUT.}
    \label{fig:trans-pC-acceptcorr}
\end{figure}

While there is a significant body of data across the full range for the total reaction cross section, the data for transparency is scarce. 
No data is shown in Fig.~\ref{fig:reac-xs-trans-pC} because a computation of acceptance correction factors to match the transparency data from electron beams is beyond this study. Fig.~\ref{fig:trans-pC-acceptcorr} shows a partial accounting of this effect. 
NuWro made transparency calculations with and without the experimental acceptance effects~\cite{Niewczas:2019fro}.
The ratio is used to estimate the impact of acceptances on the other model results by using the ratio of acceptance-corrected to Monte Carlo transparencies as an energy-dependent scale factor that is applied to all the calculations in Fig.~\ref{fig:trans-pC-acceptcorr}.  
We see that these estimated acceptance corrections put all calculations in reasonable agreement with the data~\cite{Garino:1992ca,Abbott:1997bc,ONeill:1994znv} with the exception of NuWro.
The effect of short range NN correlations has increased the transparency calculation so that it is now above the data~\cite{ONeill:1994znv,Dutta:2003yt,Rohe:2005vc}. The lowest energy points are of particular interest because of the sensitivity to nuclear effects. Although the calculations match the measurement of transparency for protons of kinetic energy $180$~MeV \cite{Garino:1992ca}, they are somewhat above the newer data points at kinetic energy $\sim$ 350 MeV~\cite{Dutta:2003yt,Rohe:2005vc}.  None of the calculations reported here show such a steep rise.

\begin{figure}[th!]
    \centering
    \includegraphics[width=0.48\textwidth]{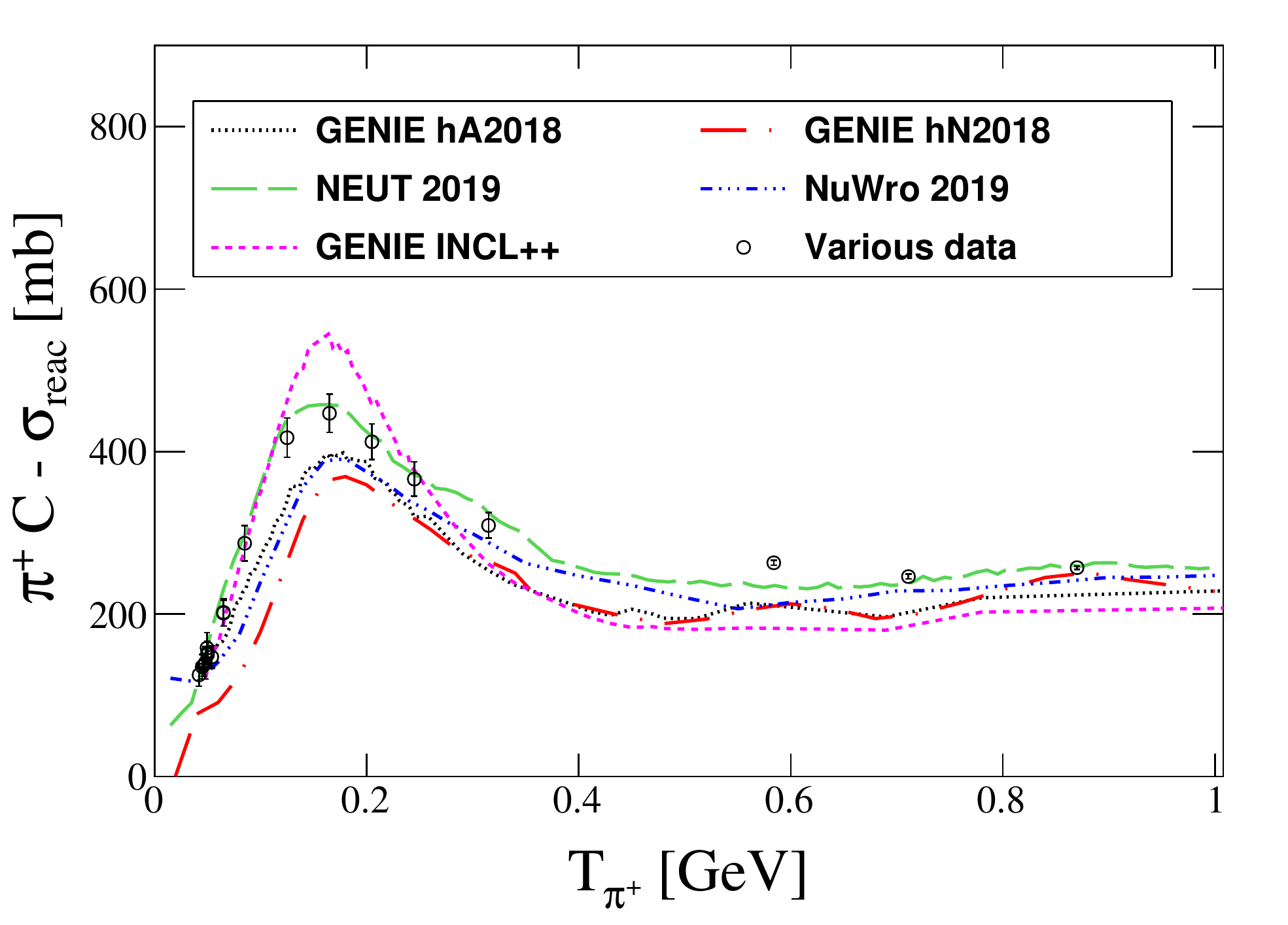}
    \includegraphics[width=0.48\textwidth]{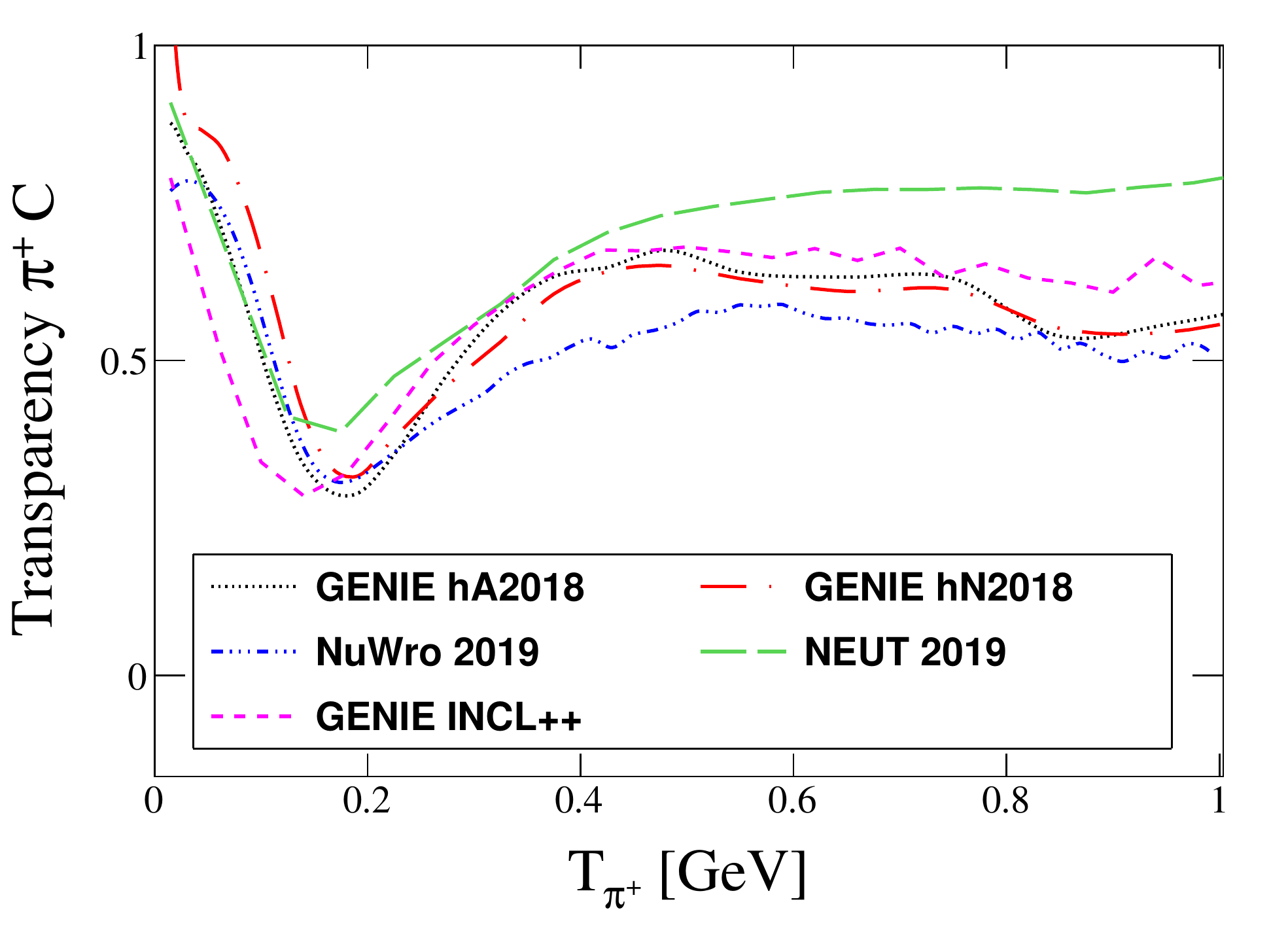}
    \caption{Total reaction cross section and transparency for $\pi^+$-carbon.  Available data~\cite{PinzonGuerra:2018rju} are shown along with calculations from GENIE, NuWro, and NEUT.  }
    \label{fig:reac-xs-trans-piC}
\end{figure}

Plots comparing \sigreac and transparency are shown for $\pi^+$ and carbon $^{12}C$ target in Fig.~\ref{fig:reac-xs-trans-piC}.
Here, the dominant feature is a peak corresponding to excitation of the $\Delta$(P$_{33}$(1232)) resonance at kinetic energy of about 165 MeV. This corresponds to a dip in the transparency results.

As was seen for protons, values of \sigreac for high energy pions (here larger than about 400 MeV) have reasonable agreement among the calculations, slightly underestimating a few existing experimental points.  However, the spread of the simulations for transparency is much larger for $\pi^+$ than for protons.  This is due to the extra effects of formation zone (NEUT) and differing treatments of the higher mass resonances (NuWro, GENIE INCL++).  
The effect of resonances at masses above the $\Delta$ is seen for GENIE {\it hA} and {\it hN}, but not for the others.  If a precise measurement could be made, these features could be tested.

Treatments of the $\Delta$ resonance in nuclei have been studied with pion~\cite{Freedman:1982yp} and electromagnetic~\cite{Drechsel:1999vh} probes.  They typically find small shifts and increases in width for nuclei.  The codes studied here have minimal corrections to the $\nu\rightarrow\Delta$ vertex and only INCL++ treats the $\Delta$ as a propagating particle.  Due to differing treatments of the interactions and nuclear models, the variations among the simulations are significant for the $\Delta$ peak in both \sigreac and transparency.  At the same time, the {\it hN}, {\it hA}, and NuWro results are close together for \sigreac and transparency for kinetic energies below roughly 300 MeV.  Since {\it hN} and  NuWro share usage of the medium corrections of Salcedo-Oset~\cite{Salcedo:1987md} and {\it hA} doesn't have that effect, this results implies that the medium corrections in FSI aren't very important.  (See Sect.~\ref{sec:modeldep} for more detail.)  It is interesting that although INCL++ is above the other simulations for \sigreac at the peak, the prediction for transparency is shifted with respect to the others. All propagating particles are in a mean field potential in INCL++~\cite{Boudard:2012wc} which depends on the kinetic energy and position.  This potential includes both nuclear and Coulomb contributions and is not in any of the other codes.  As a result, the energy of the $\pi^+$ is shifted and the dip in transparency moves to lower energy.  It is also notable that NEUT is in excellent agreement for \sigreac because the $\pi$N cross sections were fit to it~\cite{PinzonGuerra:2018rju}.  Although the NEUT result is above {\it hN}, {\it hA}, and NuWro for \sigreac, it is also above the other results for transparency.  This is due to the formation zone effect (see Sect.~\ref{sec:modeldep}). 

\begin{figure}[th!]
    \centering
    \includegraphics[width=0.48\textwidth]{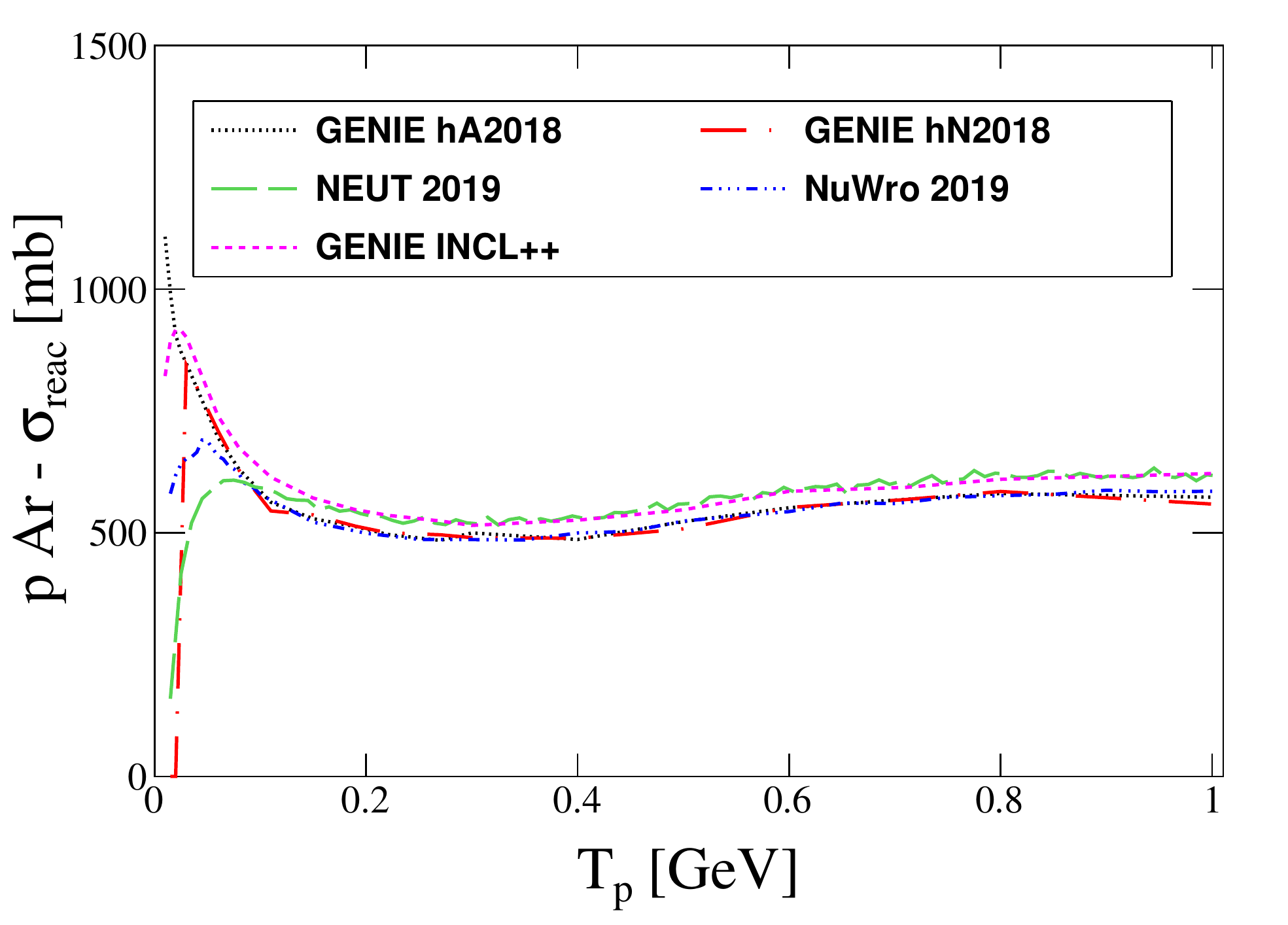}
    \includegraphics[width=0.48\textwidth]{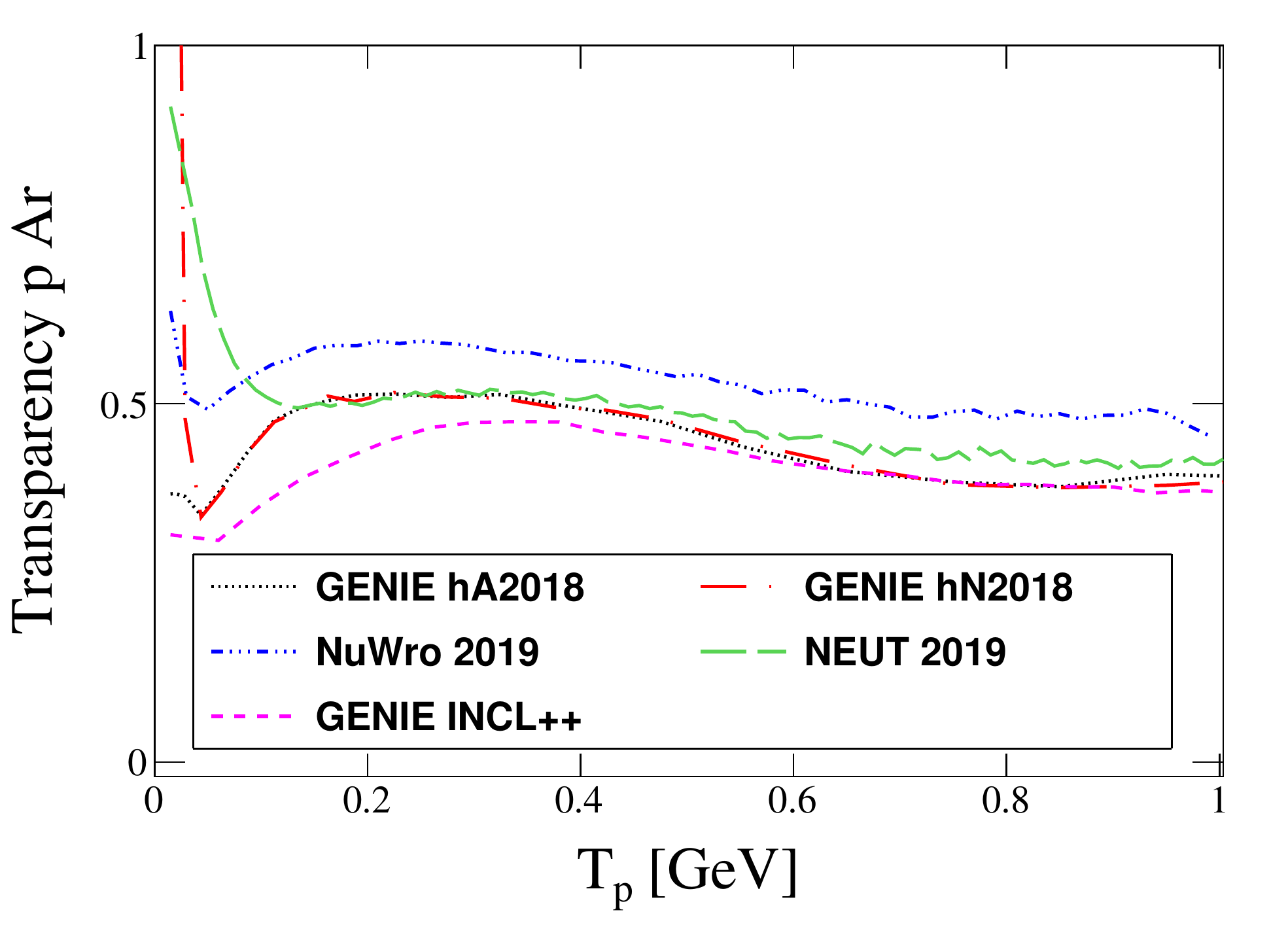}
    \caption{Subject of plots is same as in Fig.~\ref{fig:reac-xs-trans-pC} but for the proton-argon interaction.  Since there is no data available, only {\it Monte Carlo} calculations from GENIE, NuWro, and NEUT are shown.}
    \label{fig:reac-xs-trans-pAr}
\end{figure}

\begin{figure}[th!]
    \centering
    \includegraphics[width=0.48\textwidth]{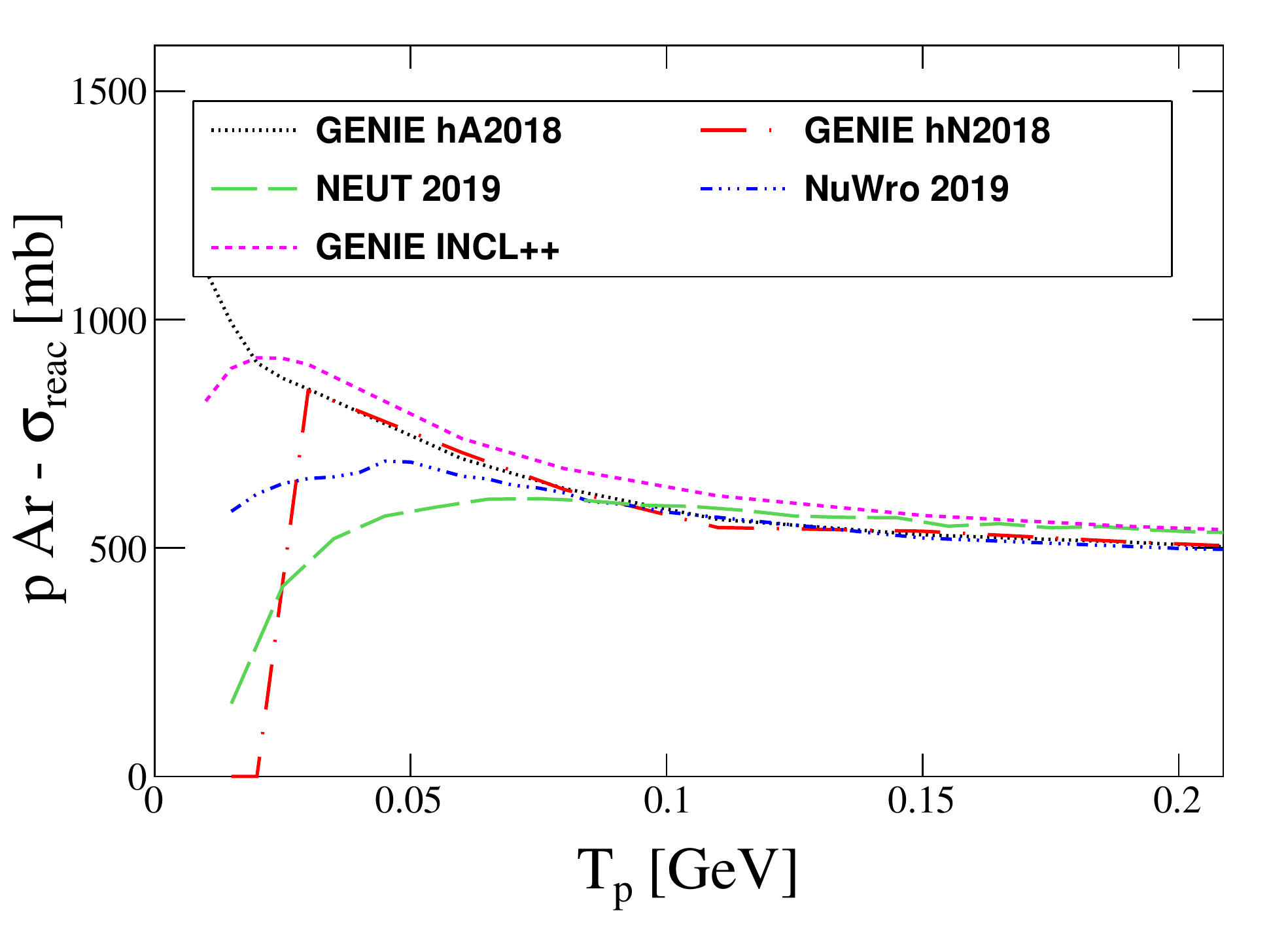}
    \includegraphics[width=0.48\textwidth]{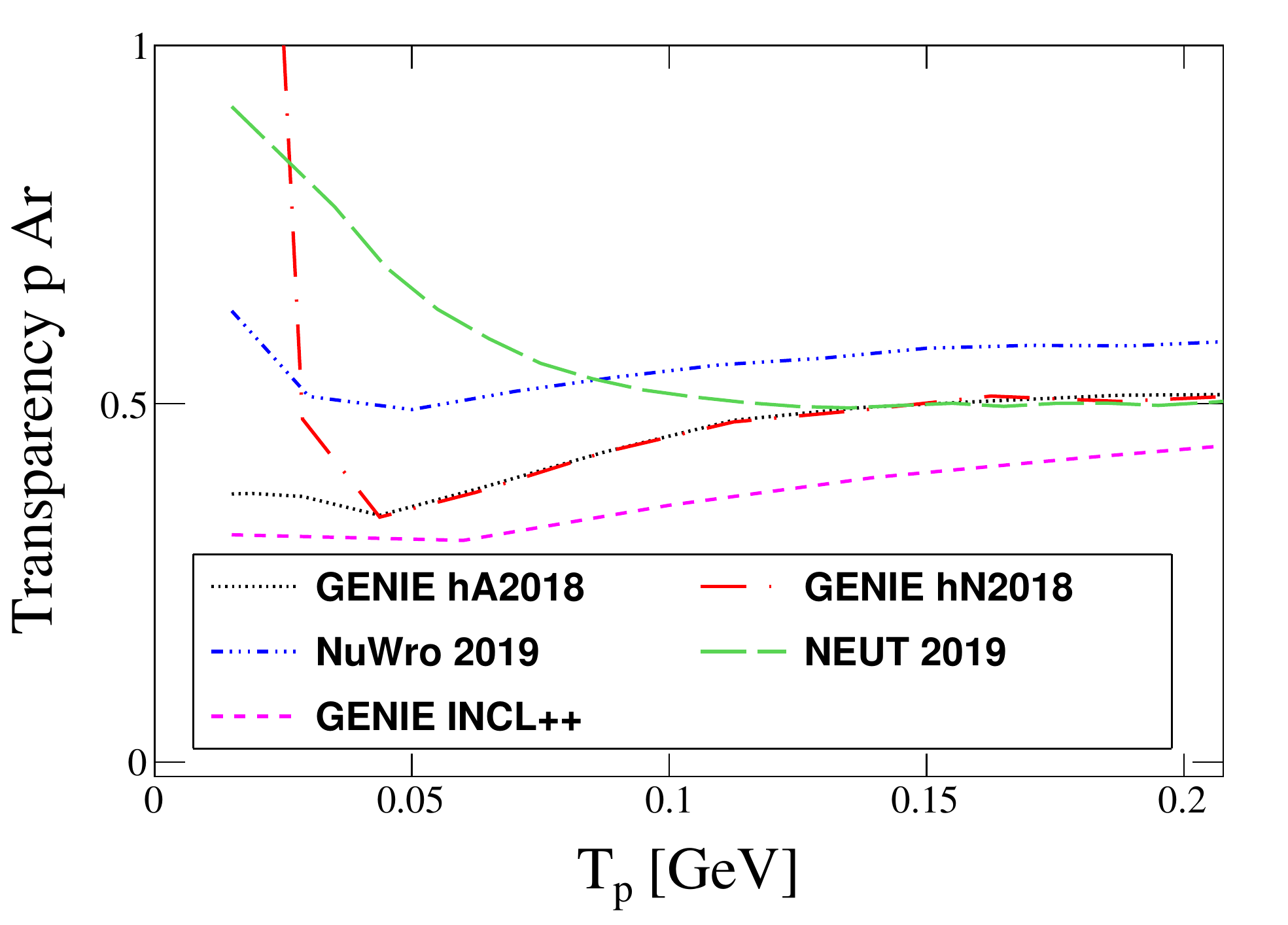}
    \caption{Subject of plots is same as in  Fig.~\ref{fig:reac-xs-trans-pAr} except for an expanded scale to show details.  Since there is no data available, only {\it Monte Carlo} calculations from GENIE, NuWro, and NEUT are shown.}
    \label{fig:reac-xs-trans-pAr-zoom}
\end{figure}

\begin{figure}[th!]
    \centering
    \includegraphics[width=0.48\textwidth]{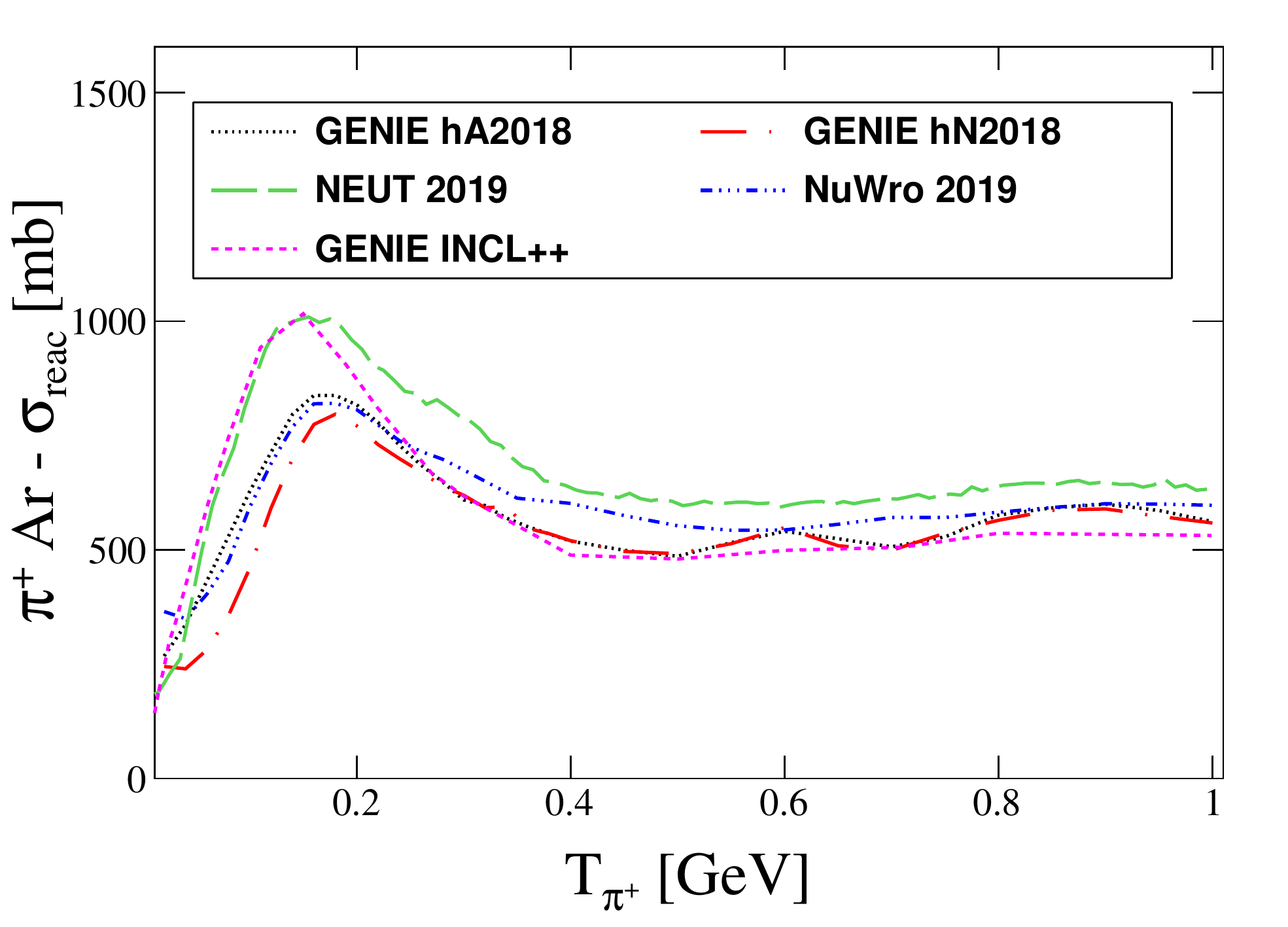}
    \includegraphics[width=0.48\textwidth]{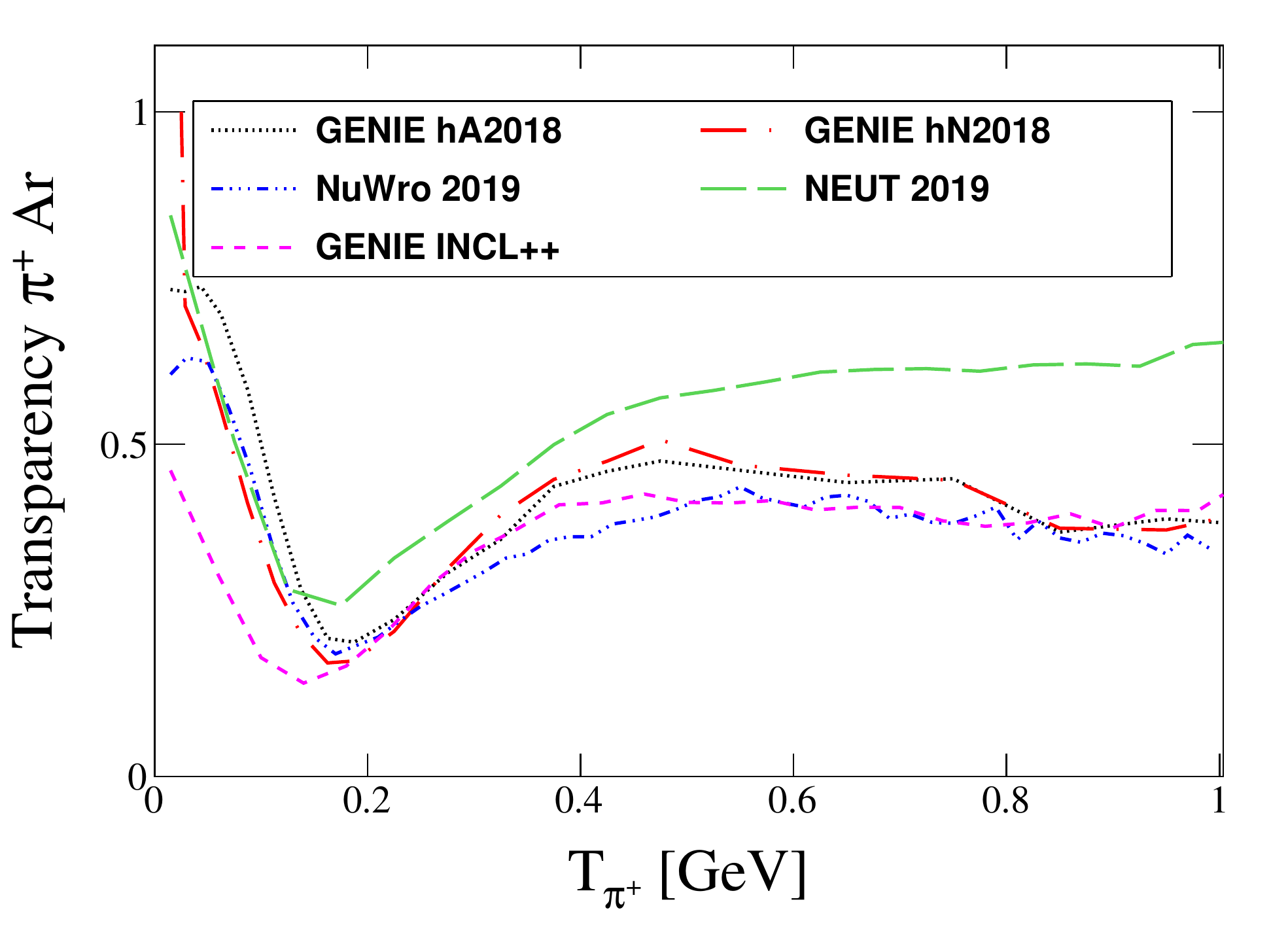}
    \caption{Total reaction cross section and transparency, same as Fig.~\ref{fig:reac-xs-trans-piC} except for $\pi^+$-argon.  Since there is no data available, only {\it Monte Carlo} calculations from GENIE, NuWro, and NEUT are shown.}
    \label{fig:reac-xs-trans-piAr}
\end{figure}

To study atomic mass (A) dependence, calculations are repeated for the argon target with results shown in  Figs.~\ref{fig:reac-xs-trans-pAr}, \ref{fig:reac-xs-trans-pAr-zoom}, and \ref{fig:reac-xs-trans-piAr}.  These calculations can be directly compared with result for a carbon target in Figs.~\ref{fig:reac-xs-trans-pC}, \ref{fig:reac-xs-trans-pC-zoom}, and \ref{fig:reac-xs-trans-piC} above.  The importance of nuclear medium effects can be expected to increase as the size of the nucleus increases.  However, the gross features of each model are unchanged with this significant increase in nuclear mass.  Although many basic nuclear effects scale linearly with A, other detailed effects such as FSI scale as $A^{2/3}$ and NN correlations can vary significantly for small changes in A.  Since reasonable agreement was obtained with iron (A=56) transparency data for NuWro in Ref.~\cite{Niewczas:2019fro}, no strong dependence on nucleus is expected.  We chose argon as a second target because of its importance in neutrino oscillation experiments.  Since there is no data for this nucleus, only a comparison among the simulations is possible.  Pauli blocking is a bigger effect for protons in argon.  This and other nuclear effects make the spread of curves somewhat more pronounced.  Medium effects make the $\Delta$ \sigreac peak wider for pions with a corresponding effect in transparency.  The tentative conclusion is that A dependence is not significant or the models fail to account properly for it.

\FloatBarrier

\section{Discussion}
\label{sec:modeldep}

It is clear that there are important differences in the FSI codes analyzed here.
In addition, there is never a guarantee that implementation has been done in the same way.
The goal of this section is to explore each code in more detail to see effects of particular components which modify reaction cross section and  transparency results.
These effects include medium corrections, nucleon-nucleon correlations, and the formation zone.

\begin{figure}[th!]
    \centering
    \includegraphics[width=0.48\textwidth]{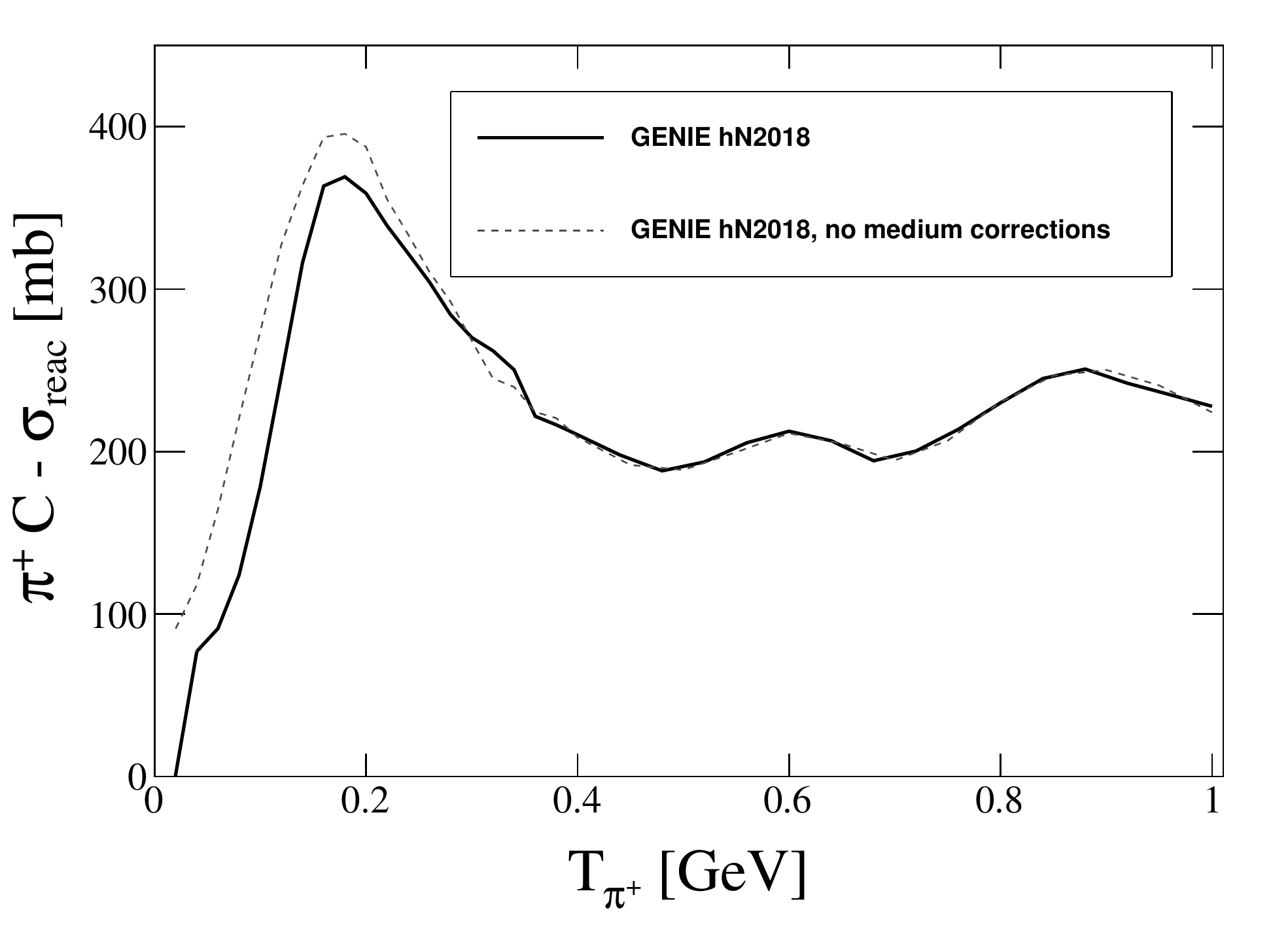}
    \includegraphics[width=0.48\textwidth]{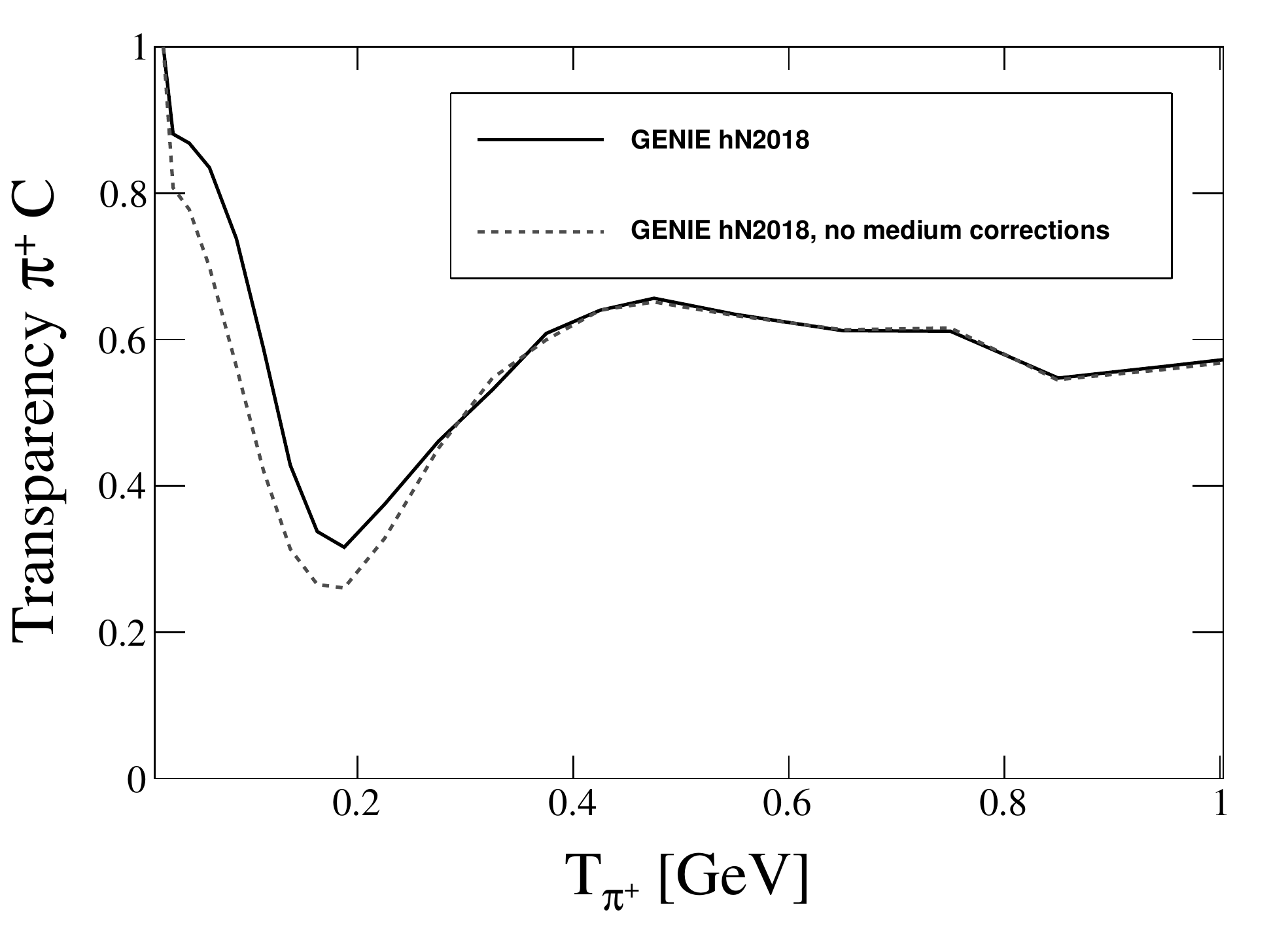}
    \caption{Total reaction cross section and transparency for pion-carbon calculated in GENIE hN2018.  This illustrates the importance of the Salcedo-Oset medium effects~\cite{Salcedo:1987md} in these quantities. "No medium corrections" means a simulation  with free $\pi$-N cross sections.  }
    \label{fig:piC-genie-model-dependence}
\end{figure}

The modular structure of GENIE allows study of many  theoretical components.  
Fig.~\ref{fig:piC-genie-model-dependence} shows the impact of the Salcedo-Oset~\cite{Salcedo:1987md} medium effects on $\pi^+$ total reaction cross section and transparency for carbon.
When these effects are taken away, no nuclear effects remain and the $\pi$N cross sections are the free values.  In fact, the structures in the results come from the underlying $\pi$N cross sections.
Although the authors give the range of model viability as 80-350 MeV as the range of applicability, GENIE also allowed it to work at pion energies smaller than 80 MeV because the effects are small there.
A small discontinuity at 350 MeV can be seen in both results.
Since there is no difference in calculation for kinetic energy larger than 350 MeV, the curves at those energies in Fig.~\ref{fig:piC-genie-model-dependence} can only differ from using samples with different random number sets or binning. 
In the GENIE adaptation, the Salcedo-Oset modification of microscopic cross section subtracts about 10\% from the total reaction cross section at the cross section peak (T$_\pi \sim$ 180 MeV) and adds about 15\% to the transparency. 

Medium effects for nucleon-nucleon interactions were explored by Pandharipande and Pieper~\cite{Pandharipande:1992zz}, also using the local density approximation.  
GENIE implemented this as a set of look-up tables as a function of nucleon energy and nuclear density for a variety of nuclei.
There is a small dependence on nucleus and that is handled with a linear interpolation between tables. 

\begin{figure}[th!]
    \centering
    \includegraphics[width=0.48\textwidth]{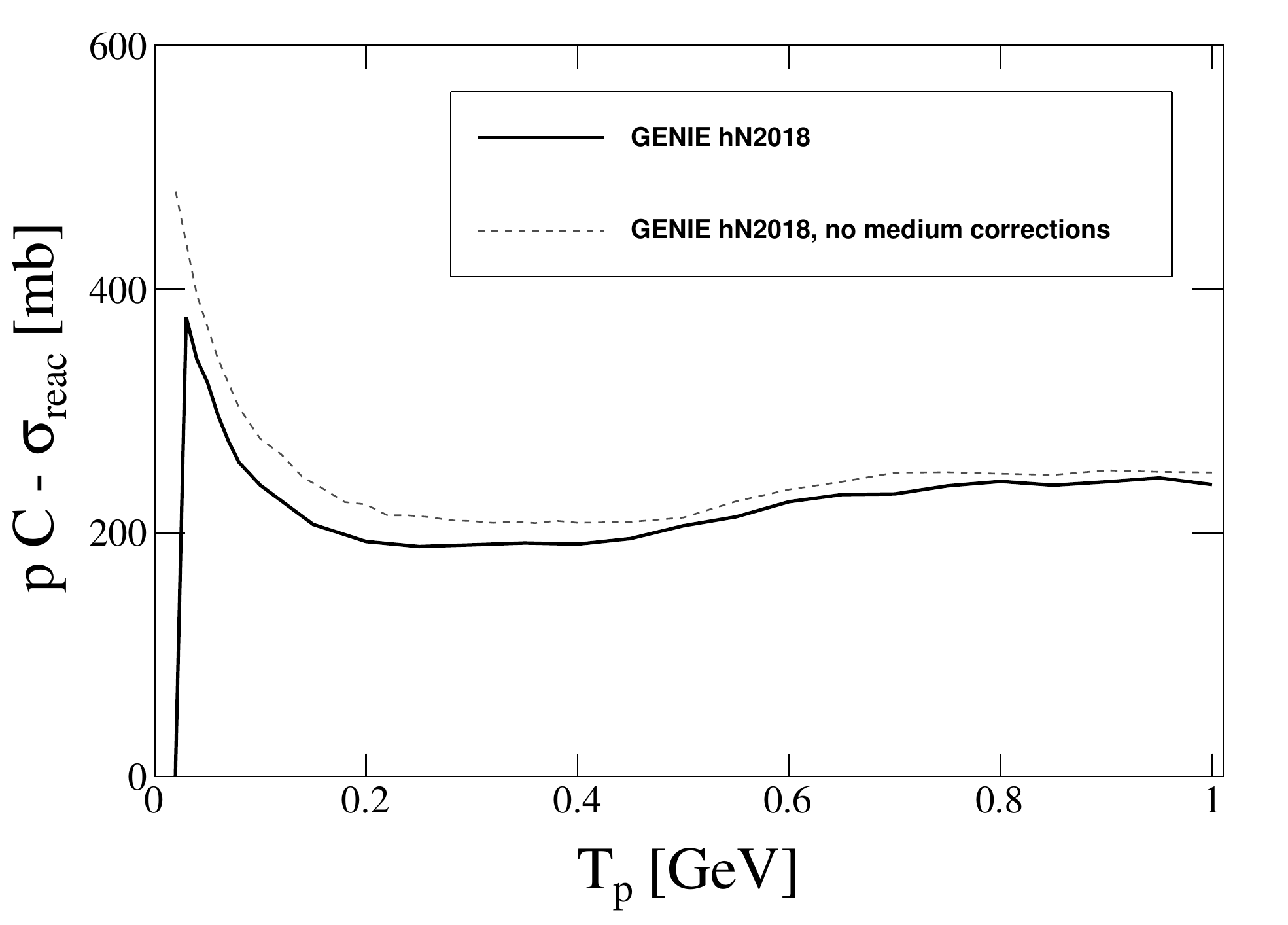}
    \includegraphics[width=0.48\textwidth]{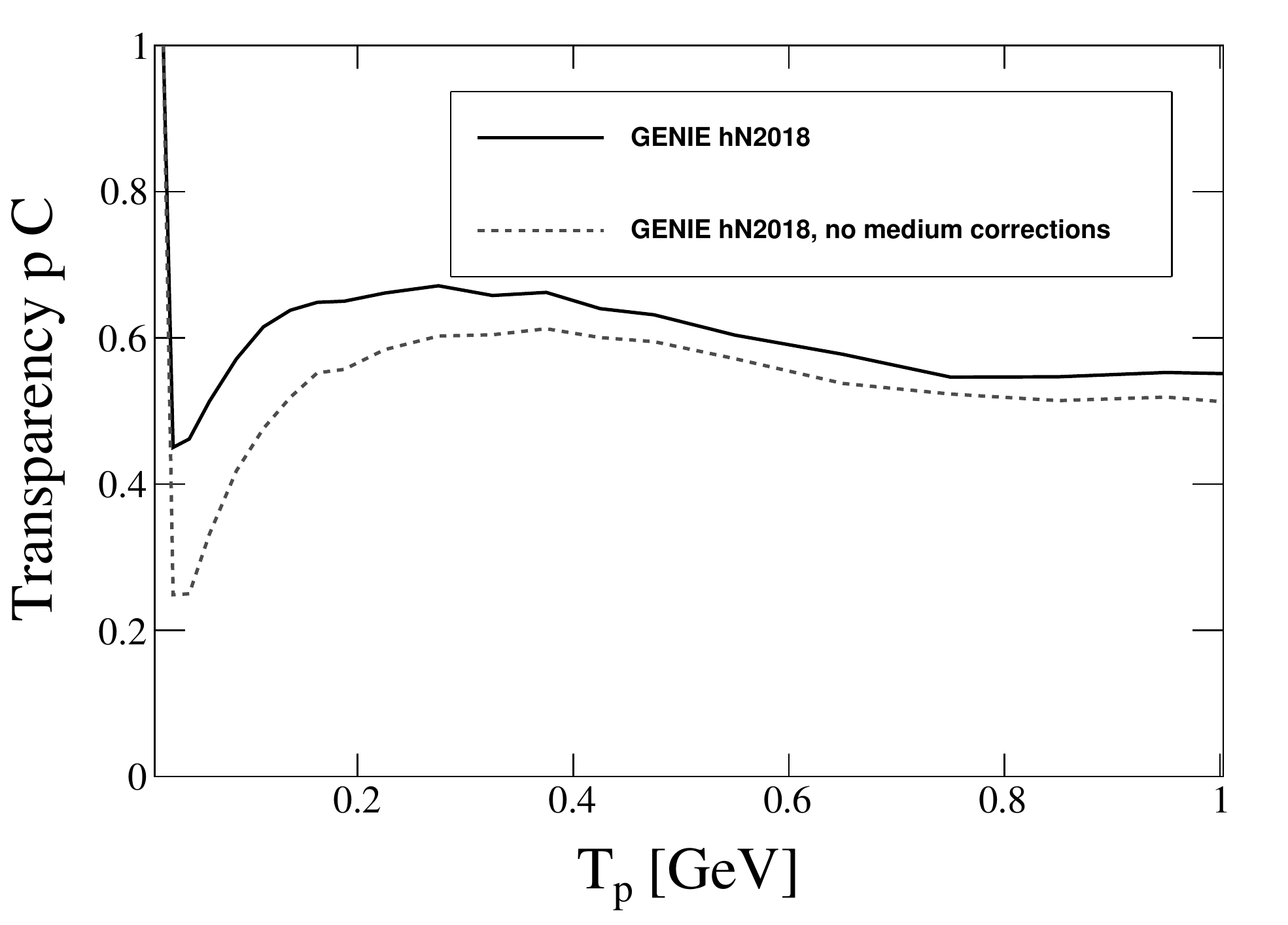}
    \caption{Reaction cross section and transparency for proton-carbon in GENIE hN2018.  Show dependence of Pandharipande-Pieper medium effects~\cite{Pandharipande:1992zz} in $\sigma_{reac}$ and transparency.  }
    \label{fig:pC-genie-model-dependence}
\end{figure}

Results for protons and carbon are shown in Fig.~\ref{fig:pC-genie-model-dependence}.
As was seen for pion interactions, the largest effect is found at lower kinetic energy where the interaction cross section is large and nuclear effects are important.
Here, the effect is a decrease in the total reaction cross section and an increase in transparency.  
The effect grows as the energy decreases, similar result was produced for NuWro in Fig.~5 in Ref.~\cite{Niewczas:2019fro}.  Here, we extend the effect to lower energy and it becomes as large as 80\%.

\begin{figure}[th!]
    \centering
    \includegraphics[width=0.48\textwidth]{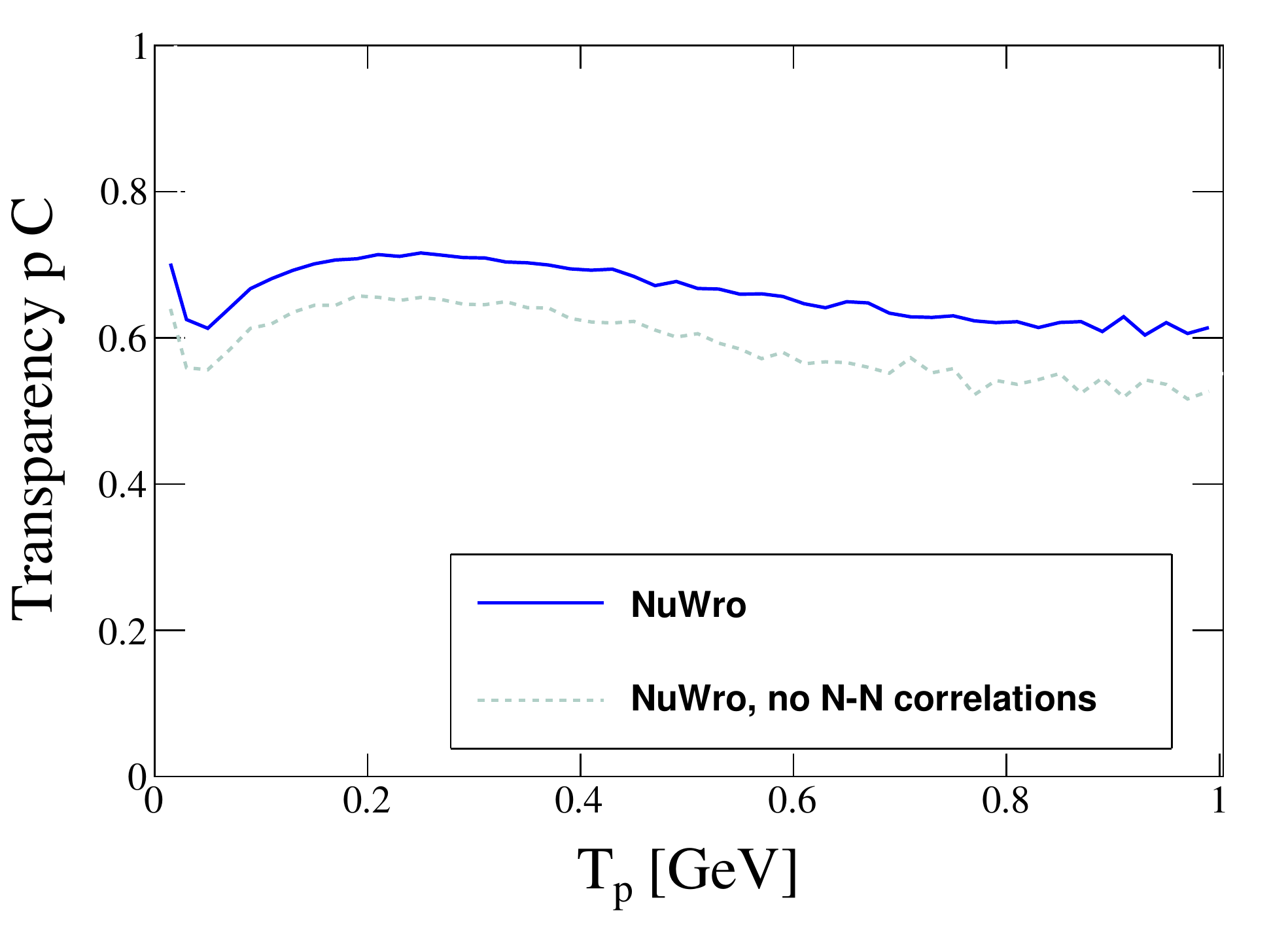}
    \caption{Transparency for proton-carbon in NuWro.  Show dependence of short-range correlation effects~\cite{Niewczas:2019fro}.  }
    \label{fig:pC-nuwro-model-dependence}
\end{figure}

In Fig.~\ref{fig:pC-nuwro-model-dependence} we show the effect of nucleon-nucleon short range correlations on the transparency results in NuWro, as described in Sec.~\ref{sect:nuwro}.  We see that the effect is to increase the transparency by 10-15\% in the whole range of proton kinetic energies. The origin of the effect is illustrated in Fig.~4 from Ref.~\cite{Isaacson:2020wlx}. Because of nucleon-nucleon correlations, the  probability of having another nucleon in a sphere of radius $\sim 0.8$~fm around any nucleon is strongly suppressed.  Interactions typically occur in the central region of nucleus with higher density. Due to correlation effects there, nucleons are more likely to leave this region avoiding any reinteraction.  Correlations do not affect reaction cross section where only the single nucleon density is relevant in the adopted approach.

\begin{figure}[th!]
    \centering
    \includegraphics[width=0.48\textwidth]{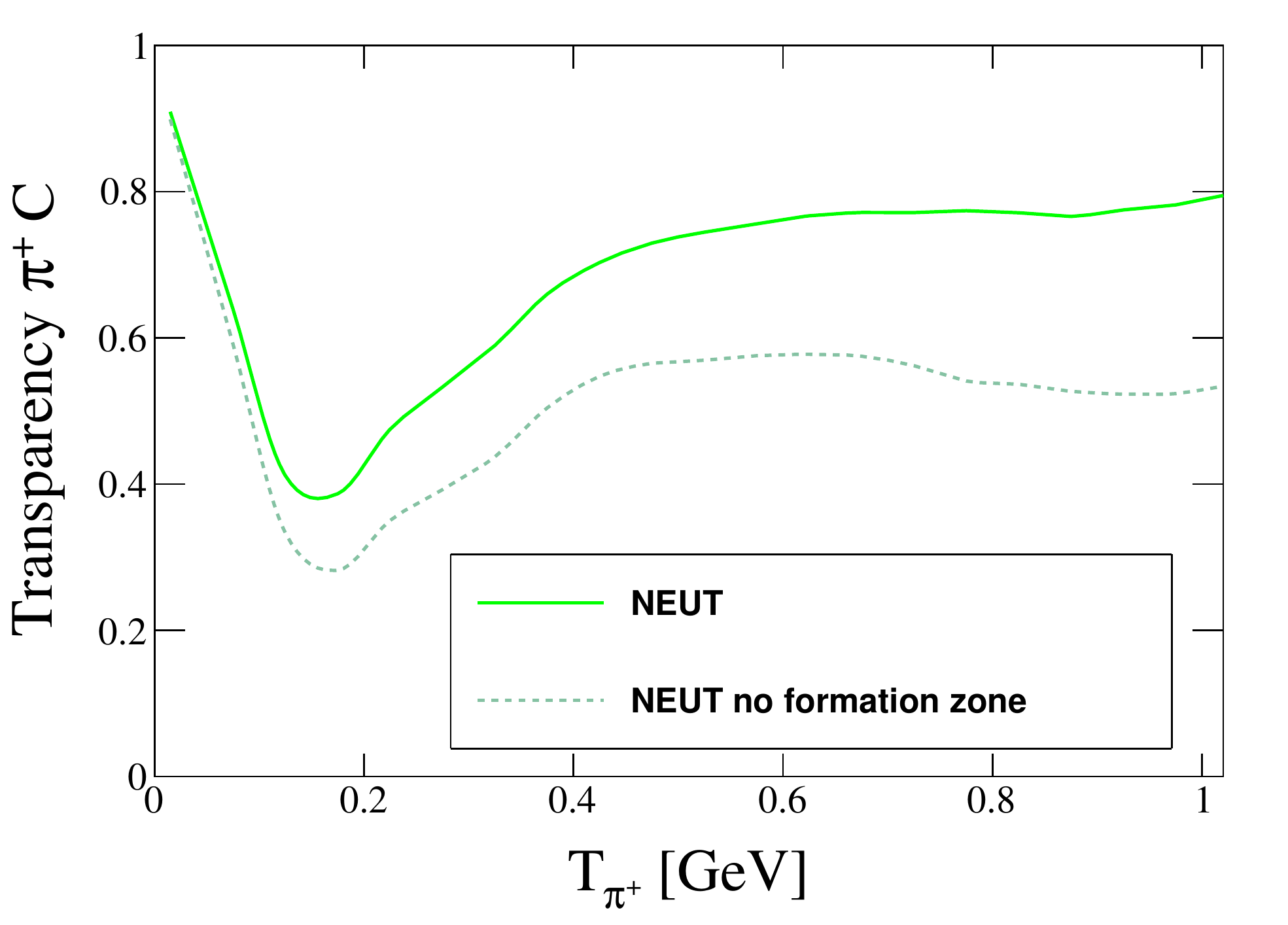}
    \caption{Effects with and without the formation zone in NEUT for $\pi^+$- carbon.  The plot shows the pion transparency in simulations of neutrino single pion production. }
    \label{fig:piC-neut-model-dependence}
\end{figure}

As described in Sect.~\ref{sec:neut}, NEUT uses a formation zone for pions produced in nuclei.  The formation zone effect is similar to that of correlations in that interactions are suppressed by giving the particle a region where it won't interact.  This effect increases the $\pi^+$ transparency for a wide range of energies as shown in Fig.~\ref{fig:piC-neut-model-dependence}.  This large effect could be easily tested in a pion electro-production experiment.

\begin{figure}
    \centering
    \includegraphics[width=0.48\textwidth]{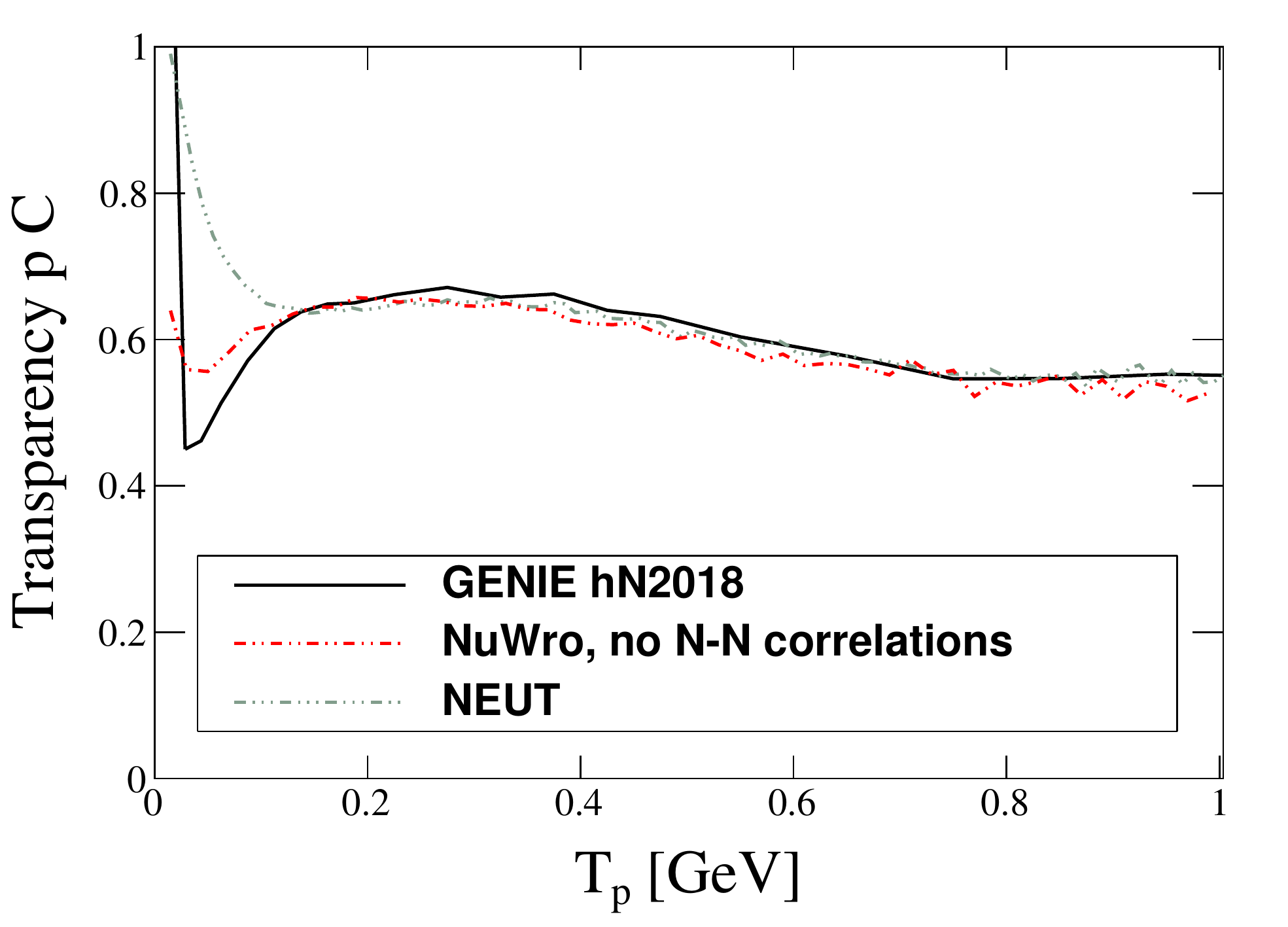}
    \caption{Comparison of generator predictions of transparency for proton-carbon.  }
    \label{fig:gen_comp_pC}
\end{figure}

\begin{figure}
    \centering
    \includegraphics[width=0.48\textwidth]{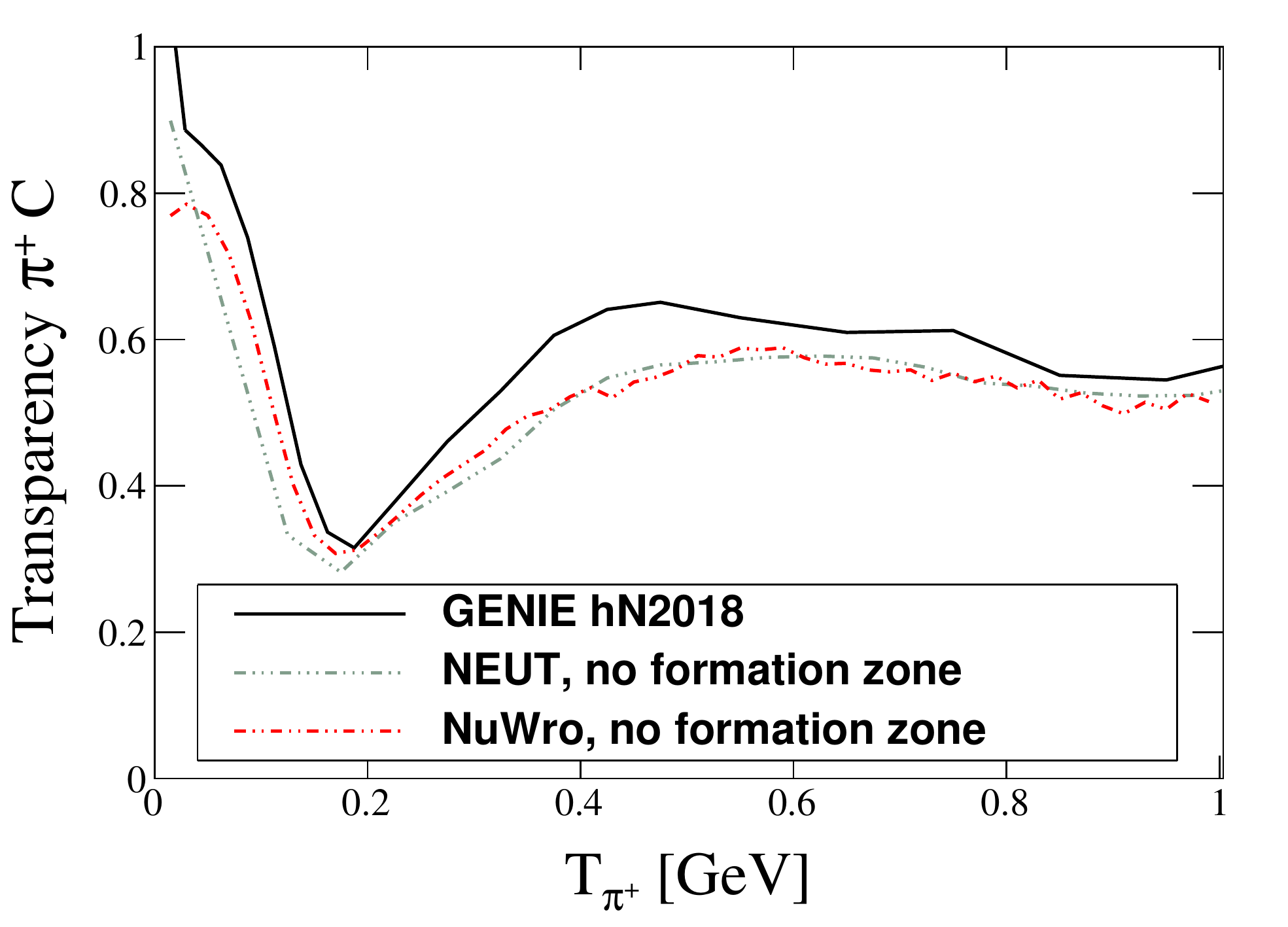}
    \caption{Comparison of generator predictions of transparency for $\pi^+$-carbon.  }
    \label{fig:gen_comp_piC}
\end{figure}

Generator codes are developed by independent groups which sometimes make different choices.  
Therefore, a direct comparison is both interesting and perhaps hard to interpret.
In Figs.~\ref{fig:gen_comp_pC} and \ref{fig:gen_comp_piC}, we attempt to make a direct comparison between the generators of transparency for proton-carbon and $\pi^+$-carbon, respectively, using the same theoretical inputs. 
For Fig.~\ref{fig:gen_comp_pC}, the NuWro NN-correlations are removed.  That result is then directly compatible with the {\it hN} result from Fig.~\ref{fig:reac-xs-trans-pC}.  These are compared with the NEUT result from Fig.~\ref{fig:reac-xs-trans-pC} which doesn't have medium corrections like the others.
The results for proton-carbon nicely continue the themes previously discussed.
Above $\sim$200 MeV, the calculations are in good agreement showing a weak model dependence there.
NuWro and GENIE {\it hN} are in agreement indicating very similar implementation of the core propagation model and medium corrections~\cite{Pandharipande:1992zz}.
However, NEUT doesn't have the medium corrections and this is shown to be a significant effect in  Fig.~\ref{fig:pC-genie-model-dependence}.
The agreement between NEUT and the others in the range of 200-400 MeV must be caused by some additional difference such as the choice of nucleon-nucleon interactions.
At low energies, the curves diverge according the approximations made. 

With the NEUT formation zone removed, the comparison for $\pi^+$-carbon in Fig.~\ref{fig:gen_comp_piC} becomes more straightforward in that all calculations have the basic propagation model together with medium modifications~\cite{Salcedo:1987md}.
However, the interpretation is not as simple as for proton-carbon because GENIE {\it hN} is not in good agreement with NuWro and NEUT.  
At kinetic energies above $\sim$400 MeV, the effects involved should come from the basic propagation model, nuclear densities and pion-nucleon cross sections.
NuWro and NEUT agree with each other and GENIE {\it hN} has a different shape and larger magnitude.
Preliminary explorations indicate that the pion-nucleon cross sections employed are different according to data base used and how pion absorption is treated.  However, a full explanation and fix is beyond the scope of this article.
On the other hand, the calculations agree on the depth of the dip due to the $\Delta$ resonance with moderate differences in the resonance width.
At the lowest energies, GENIE {\it hN} predicts a larger transparency than the others due to lack of Pauli blocking.
The discrepancies in this plot need further study.

\begin{figure}
    \centering
    \includegraphics[width=0.48\textwidth]{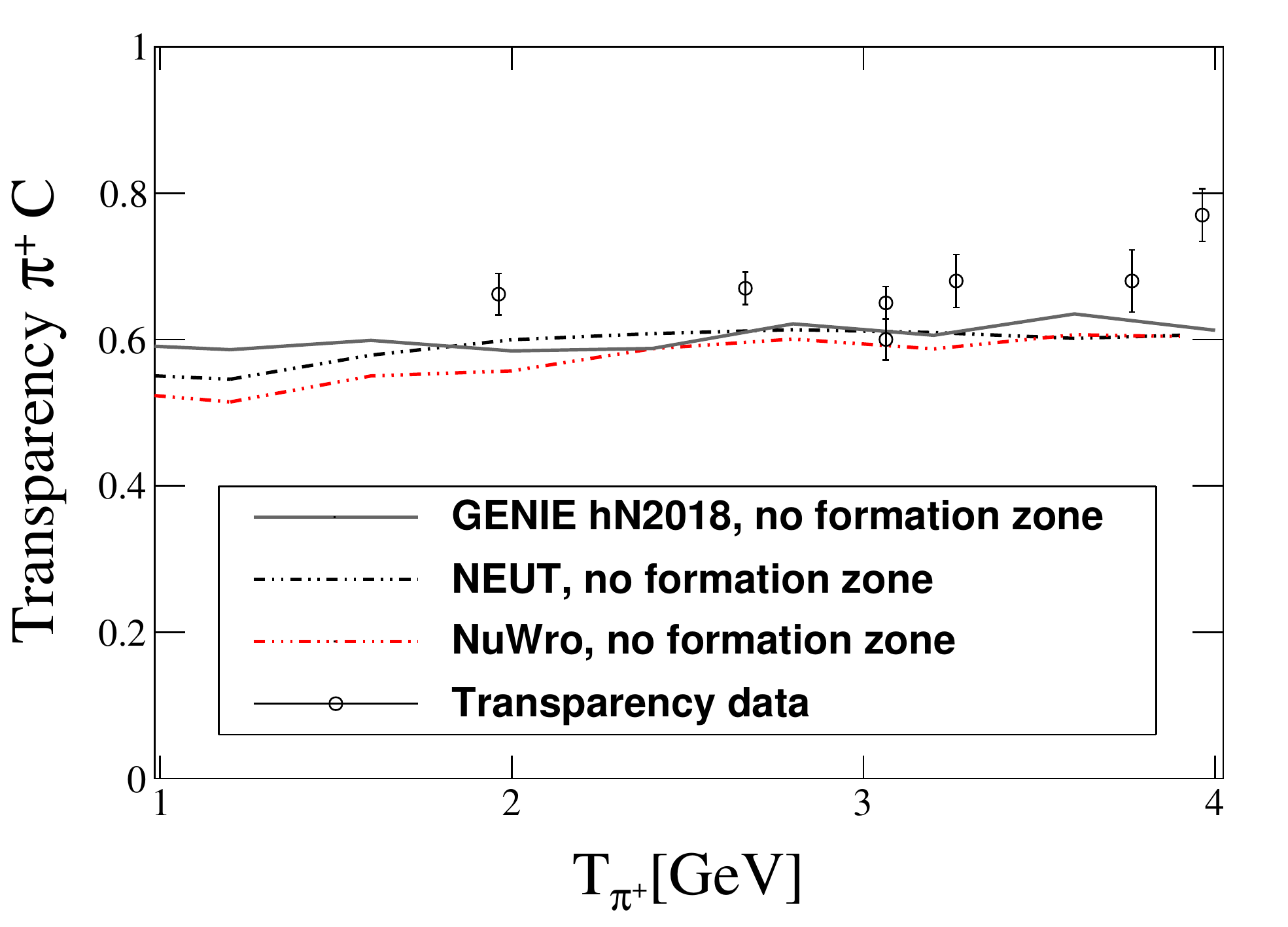}
    \caption{Comparison of generator predictions of transparency for $\pi^+$-carbon for kinetic energies above 1 GeV with data from Ref.~\cite{Qian:2009aa}.  Here, simulations using the NCDIS interaction without formation zone effects are shown.  }
    \label{fig:gen_comp_piC_NCDIS}
\end{figure}

An additional study was done to examine the disagreements in Fig.~\ref{fig:gen_comp_piC} and make a comparison
with data.  To study the transparency of higher energy pions, a new simulation was done with $\nu_e$-carbon NCDIS events. 
This gives a sample of pions up to 4 GeV.  The result for the Monte Carlo transparency is shown in Fig.~\ref{fig:gen_comp_piC_NCDIS} where the simulations are compared with data from Ref.~\cite{Qian:2009aa}.  All calculations are in qualitative agreement with the data.  Without properly accounting for experimental conditions, no detailed test can be made. 

Each curve in Fig.~\ref{fig:gen_comp_piC_NCDIS} has the standard pion formation length~\cite{Baranov:1985mb,Ammosov:2001} removed. 
Since inclusion of this
effect produced a transparency that approached 1 in this energy range, this is an
indication that the formation length derived from neutrino data~\cite{Ammosov:2001} is not needed to
describe data.  With the  models shown in Fig.~\ref{fig:gen_comp_piC_NCDIS}, the main sensitivity is to the pion-nucleon total cross sections.  Although each code has a somewhat different strategy, the results are similar.  Therefore, there is no insight into the disagreements
seen in Fig.~\ref{fig:gen_comp_piC}.

The issues in validating FSI codes are strongly dependent on probe energy.
\begin{itemize}
\item Protons (pions) above 200 (300) MeV kinetic energy have simple or no nuclear corrections as long as formation zone effects can be ignored.  Either \sigreac or transparency as a validation goal is equally correct.
\item Low energy protons (similar effects will apply to low energy neutrons) have the most sensitivity to nuclear effects.  Some of the codes in this study make severe approximations for these particles and can give untrustworthy results.
For example, Pauli blocking is significant at lower energies and its implementation varies widely among the codes.
It is notable that INCL++ has the best nuclear model and has the best agreement with proton \sigreac data.  Although all simulations have good agreement with the lowest energy transparency data at 180 MeV~\cite{Garino:1992ca}, this misses the region with the most sensitivity.  The importance of this finding will vary among experiments.  Experiments that run at low neutrino energy~\cite{Antonello:2015lea} or that need significantly improved accuracy~\cite{Abi:2020evt} will be most affected.  With the threshold for proton detection down to 47 MeV in the MicroBooNE liquid argon detector results~\cite{Abratenko:2020sga}, examination of these effects is becoming possible. 
\item Pions of any charge at energies where the $\Delta$ resonance is important are difficult for all the codes studied here.  This is well known~\cite{nustec-review,Aliaga:2020rqb} and improvements in them are necessary.  Although all codes have the effect of the propagating $\Delta$, INCL++ has the additional off-shell effects of the propagating pion which produces a noticeable effect.  Some of the codes are more than two standard deviations from the \sigreac data.  Interesting sensitivity to effects in the pion transparency await data for testing.
\end{itemize}

No significant A dependence is seen here.  The comparisons for pC (Figs.~\ref{fig:reac-xs-trans-pC} and \ref{fig:reac-xs-trans-pC-zoom}) and pAr (Figs.~\ref{fig:reac-xs-trans-pAr} and \ref{fig:reac-xs-trans-pAr-zoom}) look very similar.  This is because the quantities studied here seem to be mainly sensitive to general features that can be well-described by the models that are typically employed in these codes.  Although that suggests data for light nuclei are as good as heavier nuclei data for validating models, that could be due to the simplified nuclear treatment in the codes or the narrowness of the data studied.

The variations among the simulations are approximately equal in size for $\sigma_{reac}$ and transparency in many cases.  
For example, the fractional variation among NEUT, {\it hN}, {\it hA}, NuWro, and INCL++ results in Fig.~\ref{fig:reac-xs-trans-pC-zoom} is very similar for both quantities with protons having kinetic energy less than 200 MeV.
The same relationship can be observed for high energy hadrons and for the $\Delta$ region for pions.
In particular, the NEUT result for pion $\sigma_{reac}$ has been adjusted by tuning the underlying $\pi$N cross sections to get better agreement with data than is seen for the other codes.
This adjustment has a similar fractional effect in both \sigreac and transparency.
However, choices made in development of each code produce interesting exceptions.
One deviation to this observation comes in Fig.~\ref{fig:reac-xs-trans-piC} where NuWro is further from the GENIE calculations for transparency than \sigreac at $\sim$400 MeV pion kinetic energy.  
Another is the relationship between INCL++ and all the other calculations in Fig.~\ref{fig:reac-xs-trans-piC}.  
Although INCL++ has a larger value of \sigreac than the others, the distribution of  transparency is shifted to low $T_\pi$ due to the way INCL++ shifts the energy of the hadron in the mean field potential.  

Proper description of protons with less than 100 MeV kinetic energy requires many nuclear effects.
Codes that include Pauli blocking (NEUT, NuWro and GENIE INCL++) are qualitatively different than those that don't (GENIE {\it hA} and {\it hN}).
GENIE {\it hN} chose to stop the propagation of low energy nucleons to avoid difficulties coming from the neglect of Pauli blocking. In NEUT, events producing low energy nucleon interactions are suppressed due to Pauli blocking with different results than the other codes.
Data for low energy hadrons (both cross section and transparency, particularly for protons) should be updated and/or improved.

Medium corrections are included for pions (GENIE {\it hN} and INCL, NuWro, and NEUT according to Ref.~\cite{Salcedo:1987md}) and for protons (GENIE {\it hN} and {\it hA}, and NuWro according to Ref.~\cite{Pandharipande:1992zz}).
It is interesting that all codes chose the same models for pion and proton medium corrections.
Various comparisons have shown that at least the pion medium corrections have been implemented with an almost identical strategy for each code.
INCL++ took a different strategy in including most nuclear effects within an overall potential for all particles and gets similar results to the other codes with the notable exception of low energy nucleons and pions when the $\Delta$ excitation is important.

\FloatBarrier

\section{Conclusions}

The main motivation for this study comes from the needs of neutrino oscillation experiments. It is well established that detection of final state hadrons improves the quality of neutrino energy reconstruction
~\cite{Acero:2019ksn, Furmanski:2016wqo}. 
In addition, studies of final state hadrons allows for better understanding of dynamics of primary interactions inside nuclei.  This has an indirect effect on oscillation experiments that don't depend on hadron reconstruction but must also accurately separate events by interaction~\cite{t2k_2015}.  Both Cerenkov~\cite{t2k_2015} and scintillator experiments~\cite{Acero:2019ksn} report significant sensitivity to FSI effects.  Detection of low energy protons and pions is a high priority for liquid argon detectors~\cite{Abi:2020evt,Antonello:2015lea} for both neutrino oscillation and cross section measurements.  

$\sigma_{reac}$ and transparency provide two independent ways to validate FSI models for Monte Carlo event generators in a general way.
We have studied the relationship between them for three of the most commonly used neutrino event generators - GENIE~\cite{Andreopoulos:2009rq}, NEUT~\cite{neut}, and NuWro~\cite{Golan:2012wx}. 
While \sigreac is the total inelastic cross section as measured directly with hadron beams, transparency is the probability of escaping with no interaction when the hadron is deposited inside the nucleus which must be measured with production experiments.  
Although transparency measurements can come from either electromagnetic or neutrino beams, monoenergetic electron beams have been much more practical.
However, the application to neutrino oscillations is of great importance.
These two quantities measure different aspects of the propagation of hadrons in nuclei in a general way and 
direct comparisons show they have different dependencies on the underlying physics. 
Detailed theoretical studies of these effects are unusual~\cite{Niewczas:2019fro,Isaacson:2020wlx} and less often with codes that commonly simulate experiments.
Results for $\pi^+$ and proton propagation are studied for carbon and argon targets here.
 These targets are of great interest because many existing neutrino oscillation measurements using them are in progress or planned.  

All the codes studied here use INC approximations which are based on the idea that hadron-nucleon interactions in the medium are isolated and similar to what would happen for free nucleons.
Various effects are then employed to simulate the nuclear environment.
We find that each code implements a unique set of effects with varying effects on the results.
These codes are all in heavy use in neutrino experiments and are regularly validated against a variety of data with \sigreac being the most general data employed.
It is interesting that the nuclear effects are less  significant for \sigreac because  transparency measurements depend much more on the nuclear environment around the location of the primary interaction.
Nuclear effects such as medium corrections, short range correlations, and formation zone have measurable influences on transparency predictions.
At the same time, it must be noted that transparency measures properties closer to what is needed in neutrino oscillation experiments.

The basis for the complex simulation codes studied here is very simple -- a mean free path using nuclear density and free hadron-nucleon cross sections. 
This was explored in Sect.~\ref{sec:toy} where formulas are provided to calculate either quantity.
The main result is a single relationship between $\sigma_{reac}$ and transparency for each nucleus independent of the probe.
The connection to running codes was demonstrated through agreement with a stripped-down version of GENIE {\it hN}.
These formulas could be used as the baseline for the study of nuclear medium effects in transparency experiments.

The transparency calculations done here do not account properly for the acceptance of the measurements.
We therefore refer to them as {\it Monte Carlo transparency}.
The acceptance was taken into account in Ref.~\cite{Niewczas:2019fro} and we find that the effect is approximately a uniform enhancement of $\sim$15\% independent of energy.  
Applying this as a correction factor, all the existing codes are in reasonable agreement with the proton-carbon transparency data.

One important finding of this work is that the uncertainty in the understanding of these quantities depends very strongly on energy.
At lower energies, nuclear effects are more significant.  In fact, the picture of a localized propagating proton with a large wavelength is very questionable.
At large kinetic energy (above roughly 200 (300) MeV) for protons (pions), the mean free path is large compared to the intranuclear spacing and all the codes give similar results for both quantities.
Nuclear effects can also apply at higher energy.  For example, the addition of NN correlations in NuWro makes it $\sim$10\% larger than the others for high energy protons and the formation zone for high energy pions in NEUT produces an increased transparency for high energy pions. 

There are two kinematic regions where nuclear effects are particularly important.
In most cases, the effects come from treatments of the nuclear medium which have been imported from theoretical work describing data from a variety of probes beyond neutrinos.
These are most important at low kinetic energies and predominantly affect transparency.
Proper description of protons with less than 100 MeV kinetic energy requires many nuclear effects, including Pauli blocking, medium corrections, compound nucleus mechanisms, and Coulomb effects. 
From the comparisons here, it seems that the pN cross section attenuated by Pauli blocking is, together with medium corrections, the most important effect. 
In addition, there is significant sensitivity to nuclear effects in the way the $\Delta$ ($P_{33}$(1232)) resonance is treated.
The shift and change in width in the $\Delta$ transparency dip is different in INCL++ than the others.  Both include the effect of $\Delta$ interactions with the nuclear medium.  This effect was already seen for other probes~\cite{Freedman:1982yp} and makes the pion transparency an interesting experimental study.  An additional effect in INCL++ comes from including binding effects on the each propagating hadron.  The net effect is of opposite sign for the two methods.  

The variations among the simulations are approximately equal size for $\sigma_{reac}$ and transparency with extra uncertainty for the latter due to correlations and formation zone effects.  
That observation has important consequences for the choices in FSI model validation.
It is interesting that our results are somewhat different than the recent calculations of Isaacson, et al.~\cite{Isaacson:2020wlx}.
They have different methods to simulate interactions and a new calculation of NN correlations.
They also show a calculation which uses methods similar to what is done here.  The result is surprisingly different than what is shown in this paper.  Both of these differences need to be explored further in a dedicated study.

The uncertainty of results is expected to grow with the size of the target nucleus because the hadrons have a longer path length for interactions.  Since there is no data for argon, only the spread in Monte Carlo transparencies is available to test this hypothesis here.  There is no significant difference in the spread of values for the codes studied here between carbon (A=12) and argon (A=40).  Additional studies of heavier nuclei data with these same codes for \sigreac~\cite{PinzonGuerra:2018rju,Dytman:2011zz,Golan:2012wx} and transparency~\cite{Niewczas:2019fro} show good agreement.  Another possibility is that there is some A-dependent effect that is missing in the models presented here.  

Comparisons among event generator codes are of growing importance~\cite{tensions2016}.  In this work, a complex quantity such as transparency is studied.  One desire from such a comparison is to assess the relative accuracy of the codes.  It is clear that almost any FSI code can work for kinetic energies about $\sim$300 MeV where the nuclear corrections are small.  The treatment of low energy protons involves significant corrections due to nuclear effects which get very large for kinetic energies below $\sim$50 MeV.  Although GENIE-INCL++ has the most complete set of effects, the other codes have reasonable behavior down to $\sim$100 MeV.  Below that value, NEUT, {\it hA}, and {\it hN} diverge from the data at different values of energy.  For pions, the main sensitivity is to the handling of the $\Delta$(1232) resonance.  The differences between GENIE-INCL++ and the other GENIE models and NuWro are interesting and new experimental tests could point to improved treatment of the $\Delta$ resonance in nuclei.
Although NEUT has excellent agreement with $\sigma_{reac}$ due to tuning of the underlying $\pi$N cross sections, other codes have poorer agreement with the same data.  
Agreement of NEUT with transparency data is not guaranteed with this tuning of \sigreac.
NEUT has a significant effect due to pion formation zone that can be tested in experiments.  Even though {\it hA} is the simplest FSI model, it agrees well with the more sophisticated calculations in most situations.  

Figures~\ref{fig:gen_comp_pC} and \ref{fig:gen_comp_piC} are the closest code comparisons that can be made in this study.  Here, comparable simulations from GENIE hN2018, NEUT, and NuWro are collected for examination.  
For protons, both GENIE and NuWro have medium corrections but NEUT does not.
In general, the different treatments of low energy protons is readily apparent.
The lack of Pauli blocking in GENIE {\it hN} is seen at the lowest energies in both figures.  
However, the difference between it and the other codes for pions between 400 MeV and 1 GeV is surprising and deserves further study.  Fig.~\ref{fig:gen_comp_piC_NCDIS} extends the comparison for pion-carbon transparency to kinetic energy up to 4 GeV, testing codes in a kinematic region where nuclear effects should be less important.  All calculations are in rough agreement with data~\cite{Qian:2009aa}.  However, these calculations are made without the normally applied formation zone~\cite{Baranov:1985mb,Ammosov:2001} and disagreement with these data is significant for calculations with this effect.  This suggests that the inclusion of formation zones as implemented are inconsistent with electron scattering transparency data and should not be used.

Another important finding is that \sigreac and transparency are related in a basic way for the codes studied here, i.e. features in each quantity can be related to each other.
For example, the peak at $\sim$30 MeV in \sigreac corresponds to a dip in transparency of comparable size.
The relative importance of \sigreac and transparency as a way to validate Monte Carlo FSI codes can only be assessed in a preliminary way because of the lack of transparency data.  One good way to estimate uncertainty is to examine the range of results from reasonable codes.  We can first note that the range of results presented here is very similar for the two quantities, perhaps a little larger for transparency because of the added sensitivity to nuclear corrections.  Further, the spread of results for protons at energies above $\sim$200 MeV is much smaller than at lower energies, where nuclear effects due to Pauli blocking and medium effects cause additional uncertainties in the calculations.  Pions above $\sim$300 MeV are sensitive to formation zone effects and sensitive to the detailed treatment of the $\Delta$ resonance at lower energies.  It should be noted that the only code to accurately reproduce \sigreac over the entire energy range considered here is NEUT and that is because the $\pi$N cross sections were fit to it.

The relationship to experiment is somewhat clouded by lack of transparency data with the same quality as the \sigreac data.
At the same time, the large body of low energy proton-nucleus data is too inconsistent.
Although the transparency data for protons on carbon is significantly more sparse than for $\sigma_{reac}$, there is no data for the argon target.
Therefore, there is a strong need for more proton transparency data at all energies.
There is recent data for pion \sigreac~\cite{PinzonGuerra:2018rju}, but no data for pion transparency in the energy range studied here.  This is the most pressing need for data uncovered in this study.

At present, all FSI validations for the codes studied here use \sigreac as a primary tool.
While no significant errors are made this way as long as extra effects such as formation zone and NN correlations are externally verified, higher quality transparency data would shift emphasis toward it.
Ideally, a carefully defined balance of the two observables would be used for validation purposes.
These data provide a test of hadron interaction probability and, perhaps surprisingly, the nuclear corrections.
A full validation must include more detailed hadron interaction data such as process-dependent total cross sections~\cite{PinzonGuerra:2018rju} and inclusive double differential cross sections~\cite{Dytman:2011zz}.

Independent of the existence of data, the relationship of simulation of the two quantities can be studied.
We find that transparency is more sensitive to medium effects which have a strong effect on when a hadron first interacts when deposited into the nucleus.
Therefore, an interesting finding is that transparency measurements provide a way to study these nuclear effects in isolation through the energy dependence.

Finally, we can make suggestions for future measurements.
The quality of low energy pion and proton beam data should be improved to allow better tuning of models where the sensitivity to nuclear effects is large.
Proton transparency measurements shown here are either tagged (where the coincident electron is detected for each event~\cite{Garino:1992ca}) or untagged (relative to  calculations without the effect~\cite{ONeill:1994znv,Dutta:2003yt}).
Untagged methods are also used for studies of pion production~\cite{Dutta:2012ii} which use pions of higher energy than are examined here.
We feel that the tagged measurements have less model dependence for the energy region studied here where the quasielastic interaction is strong.
Therefore, new measurements for proton kinetic energies in this energy range similar to Ref.~\cite{Garino:1992ca} would be valuable to disentangle the nuclear effects.
The lack of any pion transparency data for kinetic energies below 1 GeV is notable.  However, a large acceptance detector such as CLAS at Jefferson laboratory could make the measurements by either method.
The dominant interaction producing a $\pi^-$ would be through a $\Delta^-$ resonance which decays to $\pi^-$n.
The tagging would require both the scattered electron and the coincident ejected neutron.
For pions of kinetic energy larger than about 350 MeV, this tagging could still be used although the resonances have a smaller decay branch to to $\pi$N.

\begin{acknowledgments}
The authors thank the US Department of Energy for support under contract DE-SC0007914.  JTS was partially supported  by  Polish Ministry of Science and Higher Education Grant DIR/WK/2017/05. YH was supported by by MEXT/JSPS KAKENHI Grant Number 18H05536. 

\end{acknowledgments}


\bibliographystyle{apsrev}
\bibliography{references}

\end{document}